\documentclass[preprint]{aastex}
\usepackage{epsf}
\def\pp{\par\parshape 2 0truecm 15.5truecm 1truecm 14.5truecm\noindent}
\newcommand{\simgt}{\lower.5ex\hbox{$\; \buildrel > \over \sim \;$}}
\newcommand{\simlt}{\lower.5ex\hbox{$\; \buildrel < \over \sim \;$}}
\newcommand{\himpc}{{\hbox {$h^{-1}$}{\rm Mpc}} }
\newcommand{\sperp}{{\scriptscriptstyle\perp}}
\newcommand{\spara}{{\scriptscriptstyle\parallel}}

\newcommand{\NL}{{\rm\scriptscriptstyle {NL}} }
\newcommand{\sigmap}{{\hbox {$\sigma_{\scriptscriptstyle {\rm P}}$}}}

\renewcommand{\P}{{\rm\scriptscriptstyle {P}} }
\newcommand{\A}{{\rm\scriptscriptstyle {A}} }
\newcommand{\CRD}{\rm\scriptscriptstyle {CRD}}
\slugcomment{\hfill RESCEU-17/99, UTAP-329/99 \\
\hfill The Astrophysical Journal {\bf 528} (2000), January 1st issue, in press}
\shorttitle{Cosmological Redshift-space distortion}
\shortauthors{ Magira, Jing and Suto}
\begin{document}

\title{
Cosmological redshift-space distortion \\
on clustering of high-redshift objects: \\
correction for nonlinear effects in power spectrum \\
and tests with N-body simulations 
}

\bigskip

\author{Hiromitsu Magira, Y. P. Jing\altaffilmark{1} 
and Yasushi Suto\altaffilmark{1}}

\affil{Department of Physics, School of Science, University of
Tokyo, Tokyo 113-0033, Japan.}

\altaffiltext{1}
{also at Research Center for the Early Universe, School of Science, 
University of Tokyo, Tokyo 113-0033, Japan.}

\received{1999 June 4}
\accepted{1999 August 16}

\begin{abstract}
  We examine the cosmological redshift-space distortion effect on the
  power spectrum of the objects at high-redshifts, which is an
  unavoidable observational contamination in general relativistic
  cosmology. In particular, we consider the nonlinear effects of
  density and velocity evolution using high-resolution N-body
  simulations in cold dark matter models.  We find that the
  theoretical modeling on the basis of the fitting formulae of
  nonlinear density and velocity fields accurately describes the
  numerical results, especially in quasi-nonlinear regimes. These
  corrections for nonlinear effects are essential in order to use the
  the cosmological redshift-space distortion as a cosmological test.
  We perform a feasibility test to derive constraints from the future
  catalogues of high-redshift quasars using our theoretical modeling
  and results of N-body simulations.  Applying the present methodology
  to the future data from on-going surveys of high-redshift galaxies
  and quasars will provide a useful tool to constrain a geometry of
  the universe.
\end{abstract}


\section{Introduction}

The observational cosmology is now entering a new and exciting phase
where the high-redshift universe can be directly and systematically
mapped by observation. Lyman-break galaxies at $z \approx 3$ (Steidel
et al. 1996, 1998) already placed a strong constraint on the nature
and evolution of the hosting haloes at the high redshifts (Jing \&
Suto 1998). The Sloan Digital Sky Survey (SDSS) will complete a
homogeneous catalogue of $\sim 10^5$ QSOs even extending $z \simgt 5$;
in fact, the highest-$z$ QSO as of this writing (SDSSp
J033829.31+002156.3 at $z=5.00$) was discovered from the SDSS
commissioning data (Fan et al.  1999). Haiman \& Loeb (1999) predict
that the {\it Chandra X-ray Observatory} will detect $\sim 100$ QSOs
at $z \simgt 5$ per its $17'\times17'$ field of view.

Those datasets will necessarily and significantly change the way of
research of the clustering of objects at high redshifts. Due to the
statistical limitation of existing catalogues of QSOs, it has been
common to infer its clustering amplitude and evolution in a
model-dependent and indirect manner (but see, e.g., La Franca,
Andreani, \& Cristiani 1998); an interesting example of this
methodology is the use of the fluctuation in the X-ray background
(e.g., Lahav, Piran,\& Treyer 1997; Treyer et al. 1998). In near
future, however, one can analyze the QSOs redshift catalogues in order
to detect the clustering at $z\sim 5$ as is routinely repeated for the
existing galaxy redshift samples at $z \simlt 0.1$.

In extracting the cosmological information from such catalogues of
objects at high redshifts, there are several observational
``contaminations'' that one has to keep in mind, including
\begin{description}
\item[linear redshift-space (velocity) distortion:] in linear theory
  of gravitational evolution of fluctuations, any density fluctuations
  induce the corresponding peculiar velocity field, which results in
  the systematic distortion of the pattern of distribution of objects
  in redshift space. The analytical expression for the linear
  distortion was first given by Kaiser (1987) and then later
  elaborated by a number of people (see Hamilton 1998 for an excellent
  review).
\item[nonlinear redshift-space (velocity) distortion:] virialized
  nonlinear objects have an isotropic and large velocity dispersion.
  This {\it finger-of-God} effect significantly suppresses the
  observed amplitude of correlation on small scales (Davis \& Peebles
  1983; Suto \& Suginohara 1991).  An empirical expression for this
  effect in the power spectrum was discussed by Peacock \& Dodds
  (1994, 1996; PD) and Cole, Fisher, \& Weinberg (1994, 1995).
\item[cosmological redshift-space (geometry) distortion:] while the
  geometry of the local universe is well approximated as Euclidean,
  the global structure of the universe should be properly described by
  a general relativistic model. In particular, the comoving separation
  of a pair of objects at $z \gg 0.1$ is not determined only by their
  observable angular and redshift separations ($\delta\theta$, and
  $\delta z$) without specifying the geometry, or equivalently the
  matter content, of the universe (i.e., the cosmological density
  parameter $\Omega_0$, and the dimensionless cosmological constant
  $\lambda_0$). This generates a nontrivial anisotropy in the
  clustering pattern of objects, particularly at $z \simgt 1$ (Alcock
  \& Paczy\'nski 1979; Matsubara \& Suto 1996; Ballinger, Peacock, \&
  Heavens 1996;  Popowski et al. 1998).
\item[cosmological light-cone effect:] all cosmological observations
  are carried out on a light-cone, the null hypersurface of an
  observer at $z=0$, and not on any constant-time hypersurface.  Thus
  clustering amplitude and shape of objects should naturally evolve
  even {\it within} the survey volume of a given observational
  catalogue. Unless restricting the objects at a narrow bin of $z$ at
  the expense of the statistical significance, the proper
  understanding of the data requires a theoretical model to take
  account of the average over the light cone which has been
  extensively discussed by our group (Nakamura, Matsubara, \& Suto
  1998; Matsubara, Suto, \& Szapudi 1997; Yamamoto \& Suto 1999;
  Nishioka \& Yamamoto 1999; Suto et al. 1999a; Yamamoto, Nishioka, \&
  Suto 1999) and others (e.g., Mataresse et al.  1997; Moscardini et
  al.  1998; de Laix \& Starkman 1998).
\item[bias:] the final difficulty in relating the clustering of any
  species of astronomical objects to that of mass fluctuations is the
  biasing (Kaiser 1984). While the first three are purely dynamical
  (or gravitational) effects, the biasing of luminous objects relative
  to mass should result from a complicated and delicate competition of
  a variety of astrophysical processes, which is almost impossible to
  describe from the first principle. Therefore a parametric
  description of the biasing on the basis of simple assumptions is
  useful at this point in understanding the generic features of the
  possible effects (e.g., Bardeen et al. 1985; Mo \& White 1996; Fry
  1996; Jing 1998; Taruya, Koyama, \& Soda 1999; Suto et al. 1999b).
\end{description}
The first two effects are already important in any shallow surveys of
objects at $z \sim 0$ (see e.g., Hatton \& Cole 1998), and the last
three effects become progressively important as the survey becomes
deeper.

While one of the primary scientific goals for the cosmological surveys
is a proper understanding of the clustering evolution of luminous
objects at high redshifts, one can probe the geometry of the universe
combining the standard theory of structure formation.  This is what we
will explore in the present paper.  Our previous work (Suto et al.
1999a) examined the two-point correlation function in cosmological
redshift space, and found that the nonlinear peculiar velocity field,
or {\it finger-of-God}, cannot be neglected even on fairly large
scales. Thus here we focus on the power spectrum, which is easier to
describe the nonlinear effects theoretically (PD; Ballinger et al.
1996).  We focus on the mass power spectrum and do not consider the
biasing explicitly below, while the same methodology is applicable if
the bias is linear and scale-independent. In reality, however, the
bias, especially for galaxies and quasars, may be too complicated to
be analytically tractable (i.e., Dekel \& Lahav 1999; Taruya, Koyama,
\& Soda 1999). This issue will be studied in detail and described
elsewhere (Magira, Jing, Taruya, \& Suto 1999).

The rest of the paper is organized as follows.  First we describe a
theoretical modeling of the power spectrum of {\it dark matter}
including the cosmological redshift-space distortion in addition to the
linear and nonlinear velocity distortions (\S 2). The theoretical model
prediction is checked and calibrated against the N-body
simulations. Next we examine the feasibility of estimating cosmological
parameters by analyzing the anisotropy in the monopole and quadrupole
moments of power spectra of objects at high redshifts with the SDSS QSO
sample specifically in mind (\S \ref{sec:feasibility}).  Finally \S 4 is
devoted to discussion and conclusions.

\section{Modeling nonlinear redshift-space distortion on power
spectrum}
\label{sec:nonlinmodel}

\subsection{Distortion due to the peculiar velocity}

The power spectrum distorted by the peculiar velocity field
(neglecting the cosmological distortion for the moment),
$P^{({\rm S})}(k;z)$, is known to be well approximated by the following
expression (Peacock \& Dodds 1994; Cole et al. 1995):
\begin{equation}
\label{eq:power_in_redshiftspace}
  P^{({\rm S})}(k_\perp,k_\parallel;z) 
  = P^{({\rm R})}(k;z)
\left[1+\beta(z) \left(\frac{k_\parallel}{k}\right)^2 \right]^2
D\left[k_\parallel\sigmap(z)\right],
\end{equation}
where $k_\perp$ and $k_\parallel$ are the comoving wavenumber
perpendicular and parallel to the line-of-sight of an observer, and
$P^{({\rm R})}(k;z)$ is the power spectrum in real space. The second factor in
the right-hand-side of equation (\ref{eq:power_in_redshiftspace})
represents the linear redshift-space distortion derived by Kaiser (1987)
adopting the distant-observer approximation and the scale-independent
linear bias, $b(z)$. Then $\beta(z)$ is defined by
\begin{eqnarray} 
\label{eq:betaz}
   \beta(z) &\equiv& {1\over b(z)} \frac{d\ln D(z)}{d\ln a} \simeq
   {1\over b(z)} \left[\Omega^{0.6}(z) + {\lambda(z) \over 70}
\left(1+ {\Omega(z) \over 2}\right) \right], \\
   \Omega(z) &=& \left[{H_0 \over H(z)}\right]^2 \, (1+z)^3 \Omega_0 ,\\
   \lambda(z) &=& \left[{H_0 \over H(z)}\right]^2 \, \lambda_0 , 
\end{eqnarray}
where $H(z)$ is the Hubble parameter at redshift $z$:
\begin{equation}
H(z) = H_0\sqrt{\Omega_0 (1 + z)^3 + 
  (1-\Omega_0-\lambda_0) (1 + z)^2 + \lambda_0} .
\end{equation}

The finger-of-God effect is modeled by the damping function
$D[k_\parallel\sigmap]$, which is the Fourier transform of the
distribution function $f_v(v_{12})$ of pairwise peculiar velocities in
real space.  It has been often assumed that $f_v(v_{12})$ is
exponential with a scale-independent pairwise velocity dispersion,
$\sigmap$.  While this exponential function has been adopted in the
literature (Cole et al.  1994, 1995; PD; Ballinger et al. 1996), its
{\it predictability} was not fully checked against the N-body
simulations; the previous studies used the value of $\sigmap$
determined {\it a priori} from the simulation data themselves. In
reality, however, the formula is useful for the present purpose only
if $\sigmap(z)$ can be computed from a given $P^{({\rm R})}(k;z)$ and a set
of cosmological parameters ($\Omega_0$, $\lambda_0$, $\sigma_8$ and
$H_0$). In next subsections, we will show that
$P^{({\rm S})}(k_\perp,k_\parallel;z)$ combined with the existing fitting
formulae for nonlinear density and velocity fields is in excellent
agreement with the results of our high-resolution N-body simulations.

\subsection{Pairwise velocity distribution function from N-body simulations}

In order to quantify various effects of redshift-space distortion and
to examine the validity of theoretical predictions, we use a series of
high-resolution N-body simulations by Jing \& Suto (1998). The
simulations assume representative cosmological models in cold dark
matter (CDM) cosmogonies (Table \ref{tab:modelparam}). Each model has
three different realizations, and employs $N=256^3$ dark matter
particles in the periodic comoving cube of the boxsize $L_{\rm box} =
300h^{-1}$Mpc.  We use the transfer function of Bardeen et al.  (1985;
BBKS) characterized by the shape parameter $\Gamma$, and the
fluctuation amplitude is normalized by using $\sigma_8$ from the
cluster abundance (Kitayama \& Suto 1997).  In the conventional CDM
models, the shape parameter $\Gamma$ is written in terms of
$\Omega_0$, $h$ and $\Omega_{\rm b}$, the baryon density parameter:
\begin{equation}
\label{eq:gamma}
\Gamma=\Omega_0 h \exp(-\Omega_{\rm b}-\sqrt{2h}\Omega_{\rm b}/\Omega_0) ,
\end{equation}
(Sugiyama 1995). In order to separate the resulting effect due to the
shape of density fluctuations from those due to the geometry and
evolution of the universe, we consider cases where $\Gamma$ is treated
as an independent parameter or given by equation (\ref{eq:gamma}) in
\S 3.  In the latter case, we assume that each simulation adopts
$h\equiv \Gamma/\Omega_0$ neglecting $\Omega_{\rm b}$ for
definiteness.

\begin{figure}
\begin{center}
   \leavevmode\epsfysize=10cm \epsfbox{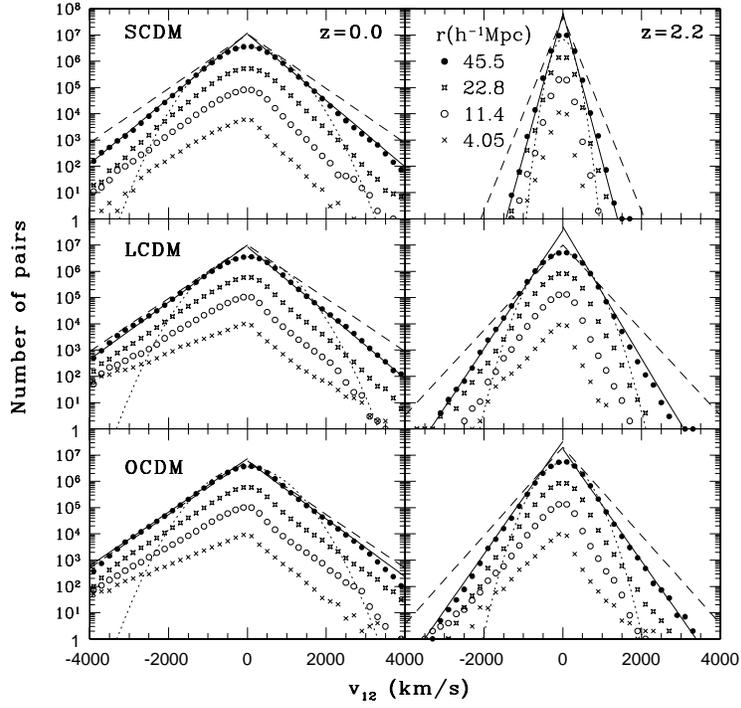}
\end{center}
\figcaption{Pairwise peculiar velocity distribution for different
  cosmological models at $z=0$ and $z=2.2$.  We compute the number of
  particles plotted is within a velocity bin of $\Delta
  v_{12}=200$km/s. Different symbols indicate the results from N-body
  simulations at different pair-separation; $r=46$ (filled circles),
  23 (stars), 11 (open circles) and 4.1 $h^{-1}$Mpc (crosses).  Solid
  lines in each panel represent the best fit to the exponential
  distribution function (eq.[\protect\ref{eq:expveldist}\protect])
  treating $\sigmap$ as a free parameter.  Dashed and dotted lines
  display the exponential and Gaussian functions, respectively, with
  $\sigmap=\sigma_{P,{\rm sim}}$.
\label{fig:pwrelvr}
}
\end{figure}

First we compute the distribution function $f_v(v_{12})$ of pairwise
velocity $v_{12}$ from the entire simulation particles in order to
determine the functional form of the damping function
$D[k_\parallel\sigmap]$.  Figure \ref{fig:pwrelvr} plots the pairwise
peculiar velocity distribution function at $z=0$ and $z=2.2$ in three
different models. Different symbols correspond to the results for
different comoving separations.  As simple analytical models for
$f_v(v_{12})$, we consider an exponential:
\begin{equation}
\label{eq:expveldist}
  f_v(v_{12}) = {1 \over \sqrt{2}\sigmap} 
         \exp\left(-{\sqrt{2}|v_{12}| \over \sigmap} \right) ,
\end{equation}
and a Gaussian:
\begin{equation}
\label{eq:gaussveldist}
  f_v(v_{12}) = {1 \over \sqrt{2}\sigmap} 
   \exp\left(- {v_{12}^2 \over 2\sigmap^2} \right) .
\end{equation}
The former is suggested to be a good approximation on nonlinear scales
both observationally (Davis \& Peebles 1983) and theoretically
(Efstathiou et al. 1988; Ueda, Itoh, \& Suto 1993; Suto 1993; Seto \&
Yokoyama 1998; Juszkiewicz et al. 1998), and the latter should be the
case if the underlying density field is purely random-Gaussian.

Solid lines in Figure \ref{fig:pwrelvr} represent the best-fit
exponential distribution for results at $r=45.5\himpc$, treating
$\sigmap$ as a free parameter. The corresponding best-fit values are
listed in Table \ref{tab:fitparam}.

Figure \ref{fig:pwrelvr} indicates that neither exponential nor
Gaussian function fits the numerical results completely. Thus it is
reasonable that the best-fit values in Table \ref{tab:fitparam} are
somewhat different from the pairwise dispersions,
$\sigma_{\scriptscriptstyle{\rm \P,sim}}$, directly evaluated from the
N-body data.  Dashed and dotted lines represent the exponential and
Gaussian curves, respectively, both of which adopt
$\sigma_{\scriptscriptstyle{\rm \P,sim}}$ as the velocity dispersion
$\sigmap$.

\begin{figure}
\begin{center}
   \leavevmode\epsfysize=5cm \epsfbox{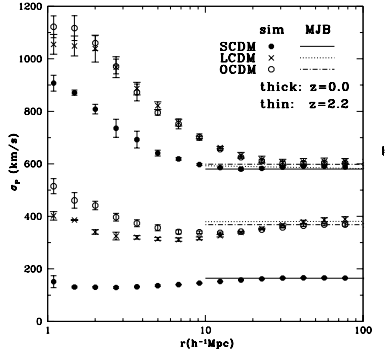}
\end{center}
\figcaption{Pairwise peculiar velocity dispersions for different
  cosmological models. Thick and thin lines correspond to the
  prediction (\protect\ref{eq:mjb}\protect) at $z=0$ and $z=2.2$,
  respectively. Different symbols correspond to the results of N-body
  simulations. Error bars represent the standard deviation among three
  realizations for each model.
\label{fig:sigmar}
}
\end{figure}

With this result in mind, however, we would still like to adopt the
above distribution functions mainly for simplicity and definiteness.
In fact we will show that this procedure is accurate enough for our
present purpose.  Thus the remaining task is to predict $\sigmap$
given a set of cosmological model parameters. In a previous paper
(Suto et al. 1999a), we find that a fitting formula by Mo, Jing, \&
B\"orner (1997; MJB hereafter) is in excellent agreement with our
numerical results.  Figure \ref{fig:sigmar} plots
$\sigma_{\scriptscriptstyle {\rm \P,sim}}$ as a function of the pair
separation $r$ for three models at $z=0$ and $2.2$. It is clear that
$\sigma_{\scriptscriptstyle{\rm \P,sim}}$ asymptotically approaches
the MJB formula:
\begin{equation}
  \label{eq:mjb}
  \sigma_{\scriptscriptstyle{\rm \P,MJB}}^2 \equiv 
\Omega(z)H_0^2
\left[1-\frac{1+z}{D^2(z)}
\int_\frac{1}{1+z}^\infty \frac{D^2(z')}{(1+z')^2}dz'\right] 
\int_0^{\infty}\frac{dk}{k}\frac{\Delta^2_\NL(k,z)}{k^2} .
\end{equation}
Note that the original expression in MJB corresponds to the proper
velocity, and the above equation (\ref{eq:mjb}) is converted in the
comoving redshift space by multiplying a factor $[(1+z)c_\perp(z)]^2 =
[(1+z)H_0/H(z)]^2$; see \S 2.3 and Suto et al. (1999a).

\begin{figure}
\begin{center}
   \leavevmode\epsfysize=7.5cm \epsfbox{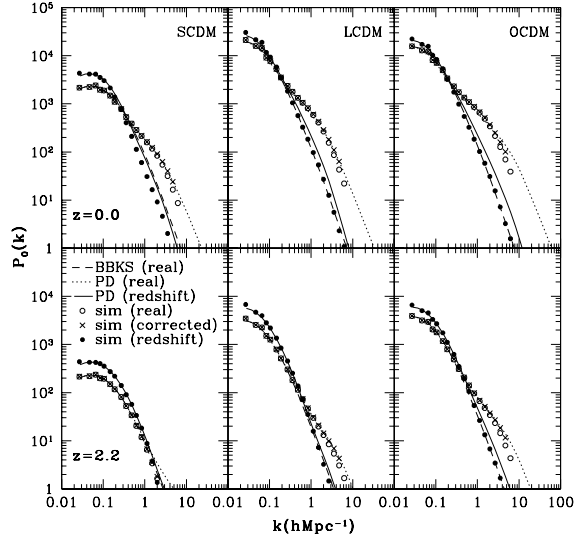}
\end{center}
\figcaption{The monopole of power spectra at $z=0$ (upper panels) and
  $z=2.2$ (lower panels) neglecting the geometrical effect in the
  redshift-space distortion. Dashed and dotted curves correspond to
  theoretical prediction based on the BBKS and PD formulae for power
  spectrum in real space, respectively. Solid curves show the
  predictions in redshift space combining the PD formula and the
  Lorentzian damping function. Different symbols indicate the results of
  our N-body simulations in real (open circles) and redshift (filled
  circles) space. The crosses refer to the power spectrum in real space
  which is corrected for the smoothing effect in the FFT.
  \label{fig:power1d} }
\end{figure}

In order to examine whether the above prescription for the theoretical
predictions is really consistent with the numerical simulations, we
compute the multipole of the power spectrum in redshift space:
\begin{equation}
\label{eq:pksldef}
  P_l^{({\rm S})}(k;z) \equiv \frac{2l+1}{2}
\int^1_{-1}d\mu P^{({\rm S})}(k,\mu;z)  L_l(\mu) ,
\end{equation}
where $\mu \equiv k_\parallel/k$ is the direction cosine, and $L_l$ is
the Legendre polynomials.  Figure \ref{fig:power1d} shows the
comparison of theoretical predictions and our numerical simulations of
the angle averaged power spectrum, $P_0^{({\rm S})}(k;z)$ at $z=0$ and
$2.2$.

Our theoretical predictions are based on the combination of the PD
formula for $\Delta^2_\NL(k,z)$ in real space and the MJB formula for
$\sigmap$. For definiteness we adopt an exponential distribution
function (\ref{eq:expveldist}) for the pairwise velocity, and the
corresponding damping function in $k$-space is a Lorentzian:
\begin{equation}
\label{eq:lorentzdampfactor}
  D[k\mu\sigmap]=\frac{1}{1+(k\mu\sigmap)^2/2} .
\end{equation}
Then equations (\ref{eq:power_in_redshiftspace}) with equation
(\ref{eq:lorentzdampfactor}) are analytically integrated; specifically
the monopole and quadrupole moments are expressed as
\begin{eqnarray}
\label{eq:pks0}
  P_0^{({\rm S})}(k,z) &=&
\left[A(\kappa)+{2\over3}\beta(z) B(\kappa)+{1\over 5}\beta^2(z) 
C(\kappa)\right]  P^{({\rm R})}(k,z) , \\
\label{eq:pks2}
  P_2^{({\rm S})}(k,z) &=&
\left[{5\over2} \left\{B(\kappa)-A(\kappa)\right\}
  +\beta(z)\left\{{4\over3}B(\kappa)+3(C(\kappa)-B(\kappa))\right\} \right. \cr
  && \left.
+\beta^2(z)\left\{{3\over2\kappa^2}(1-C(\kappa))-{1\over2}C(\kappa)\right\}
\right]  P^{({\rm R})}(k,z) , \\
  A(\kappa) &=& {1\over\kappa}{\arctan}(\kappa),
\\
  B(\kappa) &=& {3\over\kappa^2}\biggl(1-A(\kappa)\biggr),
\\
  C(\kappa) &=& {5\over3\kappa^2}\biggl(1-B(\kappa)\biggr),
\end{eqnarray}
with $\kappa(z)=k\sigmap(z)/\sqrt{2}H_0$.

Incidentally, Cole et al. (1995) derived analogous expressions for
the exponential distribution function of {\it one-dimensional} (i.e.,
not pairwise) velocity. This corresponds to a damping function:
\begin{equation}
\label{eq:coledampfactor}
 \tilde{D}[k\mu\sigma_v]=\frac{1}{[1+(k\mu\sigma_v)^2/2]^2} ,
\end{equation}
with $\sigma_v$ being the one-dimensional velocity dispersion. Then they
obtained
\begin{eqnarray}
\label{eq:colepks0}
  P_0^{({\rm S})}(k,z) &=&
\left[\tilde{A}(\tilde{\kappa})+{2\over3}\beta(z) \tilde{B}(\tilde{\kappa})
  +{1\over 5}\beta^2(z) \tilde{C}(\tilde{\kappa})\right]  P^{({\rm R})}(k,z) , \\
\label{eq:colepks2}
  P_2^{({\rm S})}(k,z) &=& \left[{5\over2} 
\left\{\tilde{B}(\tilde{\kappa})-\tilde{A}(\tilde{\kappa}) \right\}
  +\beta(z) \left\{{4\over3}\tilde{B}(\tilde{\kappa})
+3(\tilde{C}(\tilde{\kappa})-\tilde{B}(\tilde{\kappa}))\right\} \right. \cr
&& \left.  +\beta^2(z) \left\{{5\over\tilde{\kappa}^2}
(\tilde{B}(\tilde{\kappa})-\tilde{C}(\tilde{\kappa})) \right\} 
\right]  P^{({\rm R})}(k,z) , \\
\tilde{A}(\tilde{\kappa}) &=& 
    {\arctan(\tilde{\kappa}/\sqrt{2})\over 
      \sqrt{2}\tilde{\kappa}}+ {1\over 2+\tilde{\kappa}^2}, \\
\tilde{B}(\tilde{\kappa}) &=& 
     {6\over\tilde{\kappa}^2}\biggl(A(\tilde{\kappa})
      -{2\over 2+\tilde{\kappa}^2}\biggr),\\
\tilde{C}(\tilde{\kappa}) &=& 
 {-10\over\tilde{\kappa}^2}\biggl(B(\tilde{\kappa})
      -{2\over 2+\tilde{\kappa}^2}\biggr),
\end{eqnarray}
with $\tilde{\kappa}(z)=k\sigma_v(z)/H_0$. If the velocity correlation
is neglected, then $\sigma_v$ is equal to $\sigmap/\sqrt{2}$, and the
above expressions agree with our results up to
$O(\tilde{\kappa}^2)$ for $\tilde{\kappa} \ll 1$. While we use our
expression in what follows, the result is insensitive to the choice
because we mainly use the range $\tilde{\kappa} \ll 1$ for the
statistical analysis.

As Figure \ref{fig:power1d} indicates, our theoretical predictions in
redshift space (solid lines) are in good agreement with the simulation
results (filled circles) especially for $k \simlt 1 h$Mpc$^{-1}$.
For larger $k$, they start to deviate each other; while this might be
ascribed to the approximation to the velocity distribution function,
the similar discrepancy is recognized even in real space (dotted lines
and open circles). As a matter of fact, we found that this is mainly
ascribed to the smoothing effect resulting from the cloud-in-cell
interpolation in computing the Fourier transform; we used $512^3$
grids in estimating the power spectrum of simulation data.  When we
apply the correction for the smoothing effect in real space using the
method described in Jing (1992), the results (crosses) agree better
with the PD formula. Thus the discrepancy in the strongly nonlinear
regime in Figure \ref{fig:power1d} is not real.  In any case this does
not affect our conclusions in this paper because we mainly use the
range of $k \simlt 0.2 h$Mpc$^{-1}$ for the later analysis.

\subsection{Theoretical predictions of the power spectrum 
  in cosmological redshift space}

In previous subsections, we have shown using our high-resolution
N-body simulations that the power spectrum distorted by the peculiar
velocity field (without the cosmological distortion) can be predicted
fairly accurately by a combination of the existing fitting formulae.
The expression for the power spectrum in cosmological redshift space
has been derived by Ballinger et al. (1996). In this subsection, we
simply summarize their result using our own notation adopted in the
paper for definiteness. Then we examine the extent to which the
overall predictions are consistent with the N-body results in later
subsections.

Consider a pair of objects located at redshifts $z_1$ and $z_2$ whose
redshift difference $\delta z \equiv z_1-z_2$ is much less than the
mean redshift $z \equiv (z_1+z_2)/2$.  Then the {\it observable}
separations of the pair perpendicular and parallel to the
line-of-sight direction, $x_{s\sperp}$ and $x_{s\spara}$, are given as
$z\delta\theta/H_0$ and $\delta z/H_0$, respectively ($\delta\theta$
is the angular separation of the pair on the sky).  Then the mapping
of the comoving separation in real space ($x_{\sperp}$, $x_{\spara}$)
to that in CRD (cosmological redshift distortion) space is expressed
as
\begin{equation}
x_{s\sperp} (z) = x_{\perp}/c_\perp(z), \quad
x_{s\spara} (z) = x_{\parallel}/c_\parallel(z) ,
\end{equation}
where $c_\spara (z) = H_0/H(z)$, $c_\sperp (z) = H_0 (1+z)d_\A(z)/z$, 
and $d_\A(z)$ is the angular diameter distance (Matsubara \& Suto
1996; Ballinger et al. 1996; Suto et al. 1999a).

Then the power spectrum in the CRD space is
\begin{equation}
\label{eq:crdrel}
  P^{(\CRD)}(k_{s\perp},k_{s\parallel};z) 
  =\frac{1}{c_\perp(z)^2c_\parallel(z)}
P^{({\rm S})} \left(\frac{k_{s\sperp}}{c_\perp(z)},
\frac{k_{s\spara}}{c_\parallel(z)};z \right)
\end{equation}
where $k_{s\perp}$ and $k_{s\parallel}$ are the CRD wavenumber
perpendicular and parallel to the line-of-sight direction.

Substituting equation (\ref{eq:power_in_redshiftspace}), equation
(\ref{eq:crdrel}) reduces to
\begin{eqnarray}
  P^{(\CRD)}(k_{s\perp},k_{s\parallel};z) &=&
\frac{1}{c_\perp(z)^2c_\parallel(z)} 
P^{({\rm R})} \left(\sqrt{\frac{k_{s\sperp}^2}{c_\perp(z)^2}
+\frac{k_{s\spara}^2}{c_\parallel(z)^2}};z\right) \cr
  && \hspace*{-1cm}
\times \left[\frac{k_{s\sperp}^2}{c_\perp(z)^2}
+\frac{k_{s\spara}^2}{c_\parallel(z)^2} \right]^{-2} 
    \left[ \frac{k_{s\sperp}^2}{c_\perp(z)^2}+
{\beta(z)+1 \over c_\parallel(z)^2} k_{s\spara}^2 \right]^2 
    D\left[\frac{k_{s\parallel}\sigmap(z)}{c_\parallel(z)}\right].
\end{eqnarray}
Introducing
\begin{eqnarray}
k_s \equiv \sqrt{ k_{s\perp}^2 + k_{s\parallel}^2}, \quad
  \mu_s \equiv k_{s\parallel}/k_s, \quad
  \eta \equiv c_\parallel/c_\perp,
\end{eqnarray}
the above result is rewritten as
\begin{eqnarray}
\label{eq:powercrd}
  P^{(\CRD)}(k_s,\mu_s;z) 
  &=&\frac{1}{c_\perp(z)^2c_\parallel(z)}
P^{({\rm R})} \left(\frac{k_s}{c_\perp(z)}
\sqrt{1+[{1\over \eta(z)^2}-1]\mu_s^2} ; z \right) \cr
  && \hspace*{-2cm}
\times \left[1+ \left({1\over \eta(z)^2}-1\right)\mu_s^2
                \right]^{-2}
\left[1+ \left({1+\beta(z) \over \eta(z)^2}-1\right)\mu_s^2\right]^2 
~ \left[1+ \frac{k_s^2\mu_s^2\sigmap^2}{2c^2_\parallel(z)} \right]^{-1},
\end{eqnarray}
where we adopt equation (\ref{eq:lorentzdampfactor}) for the damping
function.

\subsection{Comparison of the predicted anisotropy \protect \\ \protect 
of the power spectrum with N-body simulations}

In order to compute the power spectrum in CRD space, we employ the
distant-observer approximation, calculate the power spectrum in
comoving space, and finally transform it to that in the observed frame
according to equation (\ref{eq:crdrel}). We adopted this indirect
procedure since we are mainly interested in the anisotropies in the
power spectrum. As we will show in \S 3, however, even the
angle-averaged power spectrum alone is useful as a cosmological test.
Therefore in practice one would not have to take this route, but
estimate the power spectrum directly in CRD space, because the
distribution of objects obtained by observations has already been in
the observed frame. We made sure that the estimate of the 
angle-averaged power spectrum from simulation data is almost identical
even if we first transform the simulation volume into the CRD space
and then compute the power spectrum.

\begin{figure}
\begin{center}
   \leavevmode\epsfysize=7.5cm \epsfbox{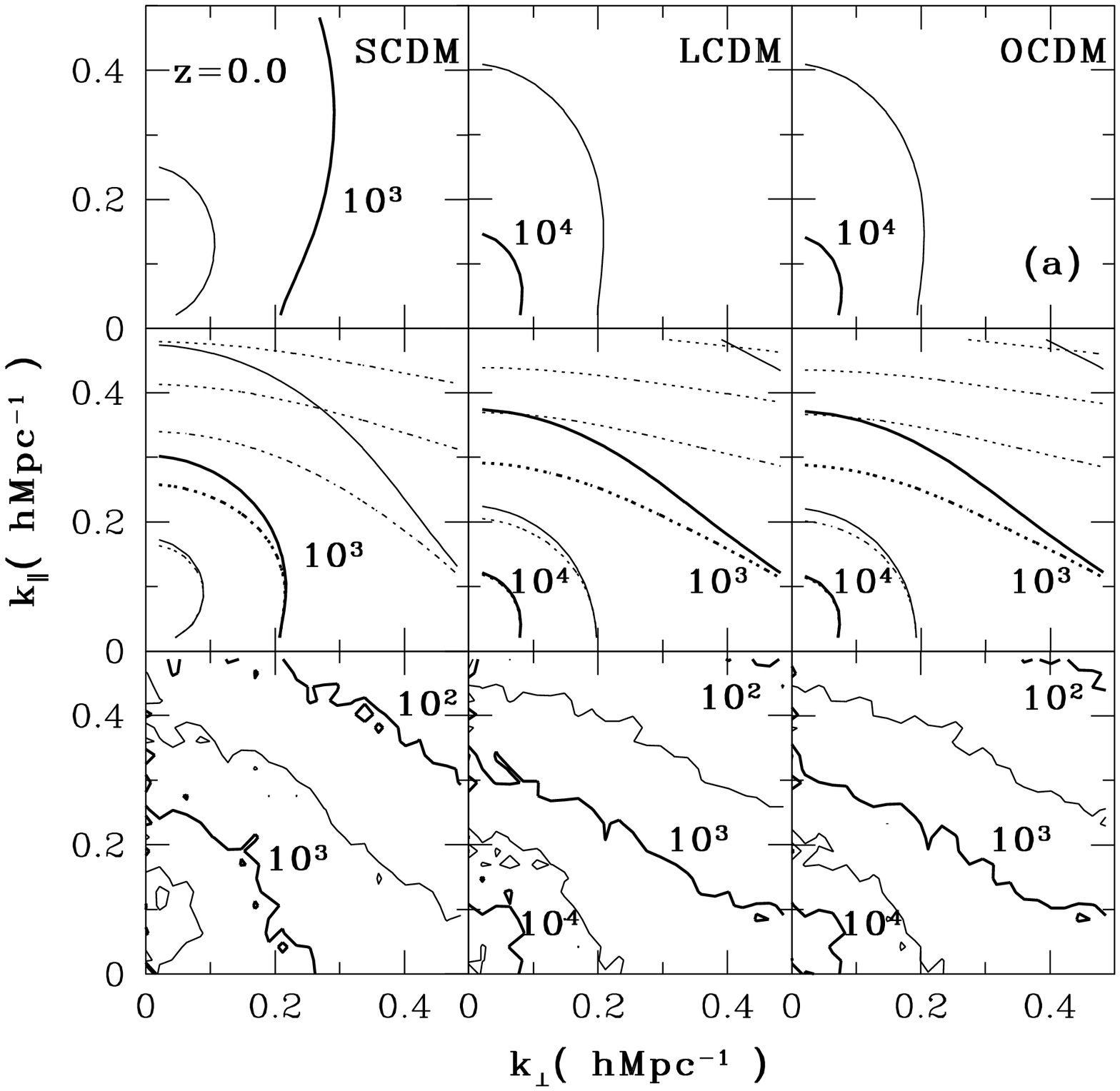}
\end{center}
\begin{center}
   \leavevmode\epsfysize=7.5cm \epsfbox{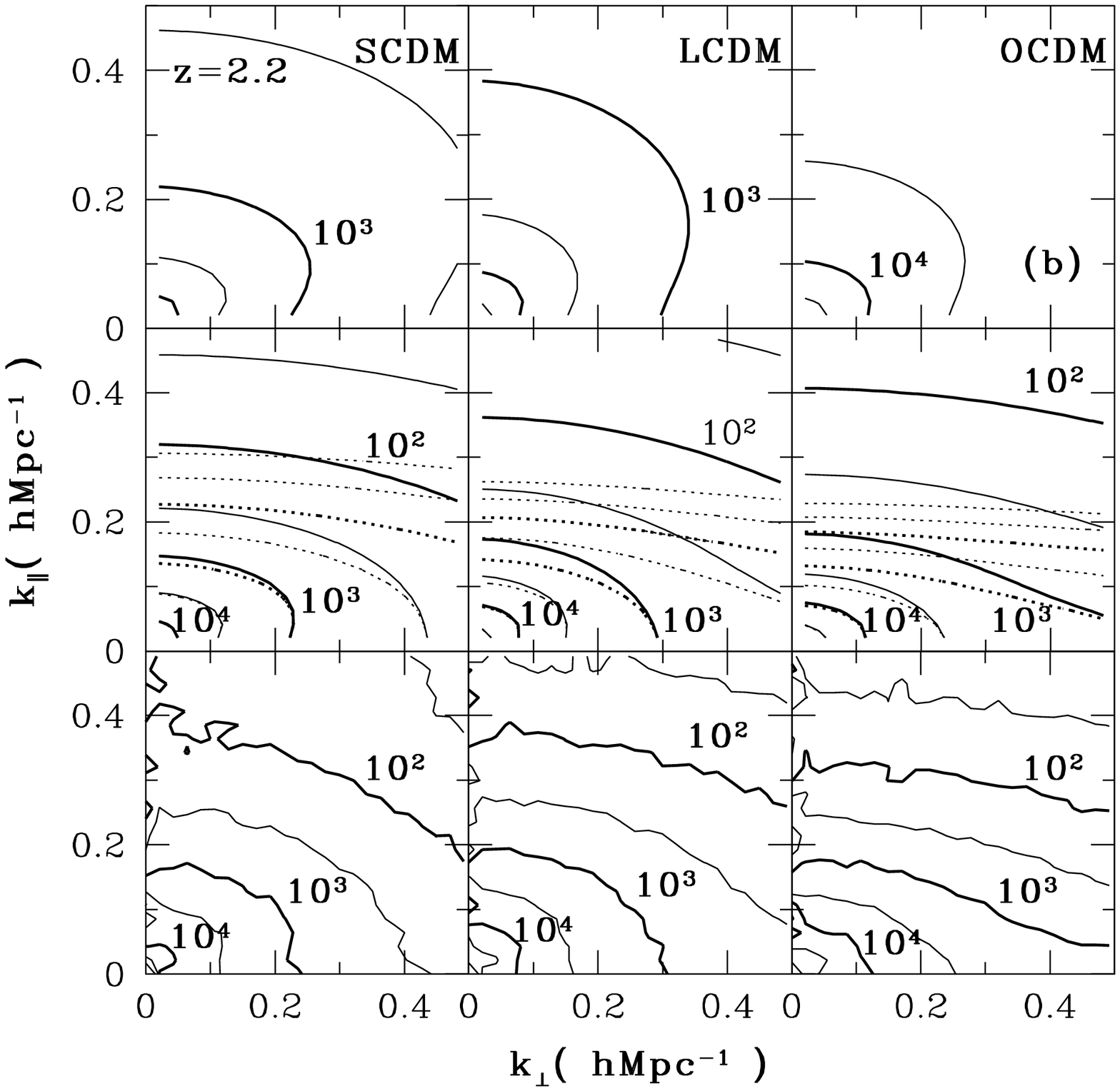}
\end{center}
\figcaption{Two-dimensional power spectra in cosmological redshift
  space; (a) $z=0$, (b) $z=2.2$.  In each figure, the upper panels
  display the predictions with the PD mass power spectrum and linear
  velocity distortion. The middle panels display the predictions with
  the PD mass power spectrum and non-linear velocity distortion. The
  Lorentzian (solid) and Gaussian (dotted) are adopted for the damping
  function describing the nonlinear velocity distortion (with
  $\sigmap=\sigma_{\scriptscriptstyle{\rm P,MJB}}$). The bottom panels display the power
  spectrum calculated from N-body simulations with all the particles
  ($N=256^3$).  \label{fig:power2d} }
\end{figure}

In Figures \ref{fig:power2d}a and \ref{fig:power2d}b, we show the
contours of power spectrum in CRD space at $z=0$ and $2.2$,
respectively.  In each figure, top panels correspond to theoretical
predictions taking account of the linear velocity distortion alone but
with the PD formulae for the power spectrum.  Middle panels
include the correction for the nonlinear finger-of-God using the MJB
formulae in equation (\ref{eq:powercrd}), and bottom panels plot the
results from N-body simulations.  As Cole et al.(1994) emphasized, the
finger-of-God effect is appreciable even at $k_s \sim 0.2 h$Mpc$^{-1}$
where the nonlinearity in {\it density field} is rather small.

In middle panels, solid and dotted contours refer to the the
prediction using a Lorentzian and a Gaussian for the damping function,
respectively.  Despite the fact that the exponential does not
perfectly match the distribution function of pairwise relative
peculiar velocity derived from simulations, the agreement of theory
and simulation in Figure \ref{fig:power2d} is rather well in this
quasi-nonlinear regime. Nevertheless the direct analysis using the
contour curves is still sensitive to the statistical noise in the
data, and we instead are forced to use the multipole expansion as in
our previous paper (Suto et al.  1999a):
\begin{equation}
\label{eq:lthmoment}
  P_l^{(\CRD)}(k_s;z) \equiv 
\frac{2l+1}{2}\int^1_{-1}d\mu_s P^{(\CRD)}(k_s,\mu_s;z) L_l(\mu_s) .
\end{equation}

\begin{figure}
\begin{center}
   \leavevmode\epsfysize=7.5cm \epsfbox{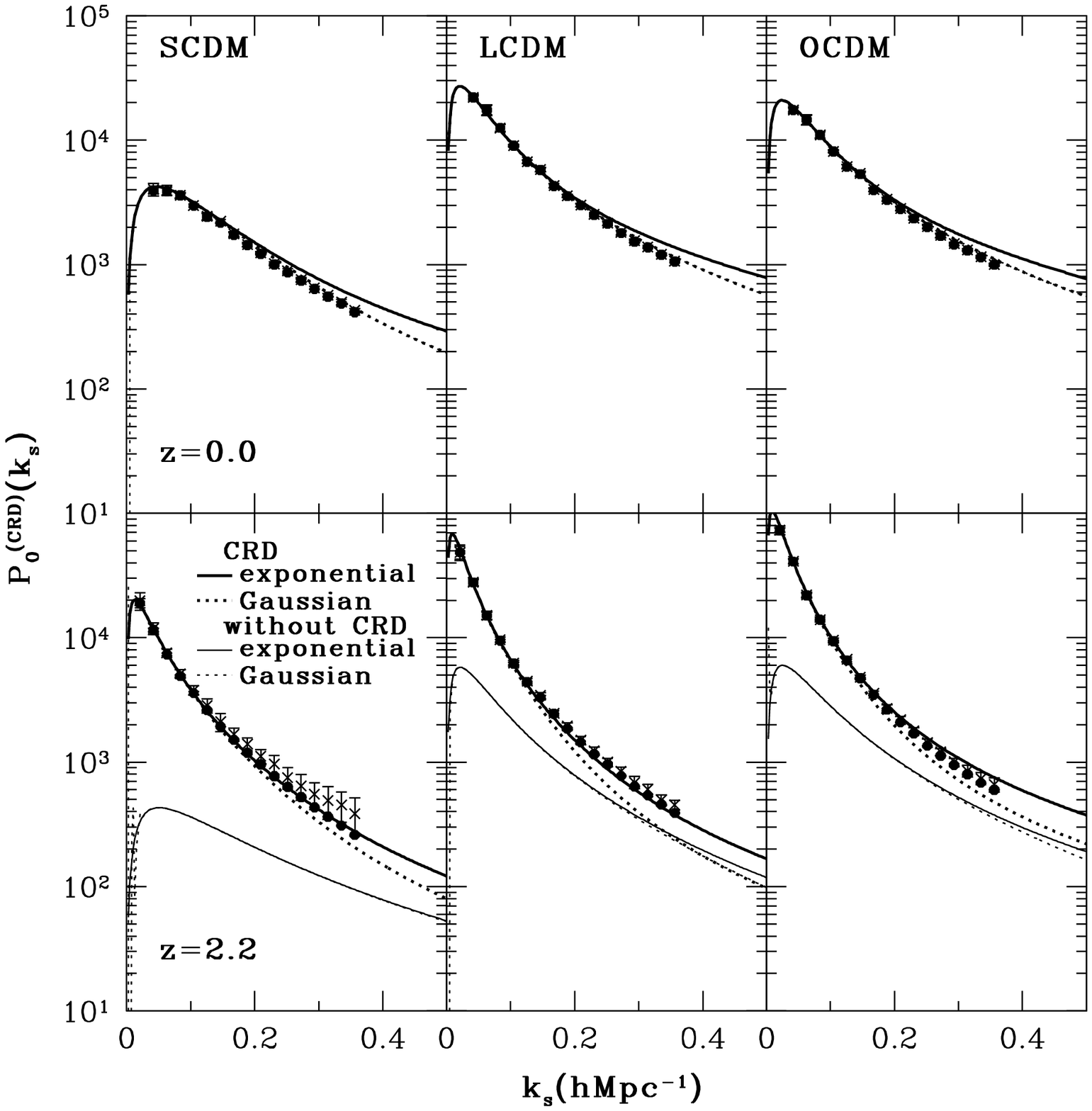}
\end{center}
\figcaption{The monopole of power spectra for different cosmologies at
  $z=0$ (upper panels) and $z=2.2$ (lower panels).  Filled circles and
  crosses correspond to the results from N-body simulation, which are
  evaluated from all particles ($1.7\times10^7$) and selected
  particles($5\times10^4$), respectively. Thick curves correspond to
  the predictions including cosmological redshift distortion effect
  with an exponential (solid) and Gaussian (dotted) damping function,
  respectively. For reference, the predictions neglecting the
  geometrical effect at $z=2.2$ are shown in thin
  curves(eq.[\protect\ref{eq:pks0}\protect]).  \label{fig:p0combcdf} }
\end{figure}

\begin{figure}
\begin{center}
   \leavevmode\epsfysize=7.5cm \epsfbox{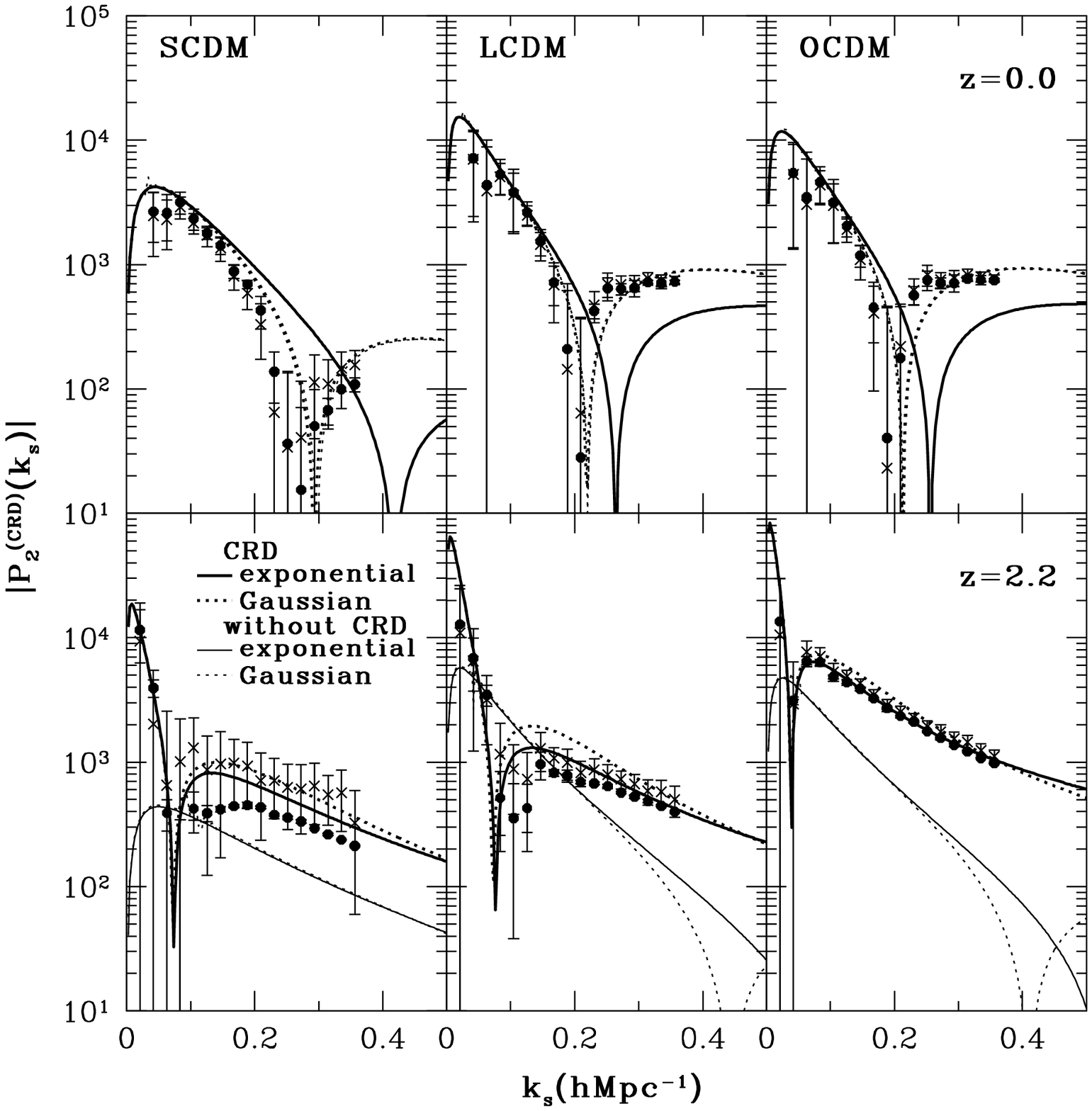}
\end{center}
\figcaption{The same as Figure \protect\ref{fig:p0combcdf} but for the
  quadrupole moment of the power spectra.
\label{fig:p2combcdf}
}
\end{figure}

Figures \ref{fig:p0combcdf} and \ref{fig:p2combcdf} plot
$P_0^{(\CRD)}(k_s)$ and $P_2^{(\CRD)}(k_s)$, the monopole and quadrupole
moments of the power spectra, at $z=0$ and 2.2.  Filled
circles and crosses represent the simulation results with all particles
($1.7\times10^7$) and randomly selected particles ($5\times10^4$),
respectively. The quoted error bars refer to the dispersion among
three different realizations in the case of all particles, and among
24 subsamples in total (i.e., 8 randomly selected subsamples for each
realization) in the case of $5\times10^4$ particles.  Note that the
major difference in amplitude between the results with and without the
geometrical effect in Figure \ref{fig:p0combcdf} comes from the
overall normalization factor $c_\perp(z)^2c_\parallel(z)$ in equation
(\ref{eq:crdrel}). This is, however, a matter of definition to some
extent, and what is important here is the difference among the
cosmological models which is much smaller than the factor
$c_\perp(z)^2c_\parallel(z)$ (Ballinger et al. 1996; Matsubara \& Suto
1996).

Numerical integration of equations (\ref{eq:powercrd}) and
(\ref{eq:lthmoment}) adopting the PD and MJB fitting formulae yields
the corresponding theoretical predictions, which are plotted in thick
solid and dotted lines for the exponential and Gaussian velocity
distribution functions.  Analytical results neglecting the
cosmological distortion (eq.[\ref{eq:pks0}]) are shown in thin lines
for reference.  In any case the difference due to the modeling is
negligible at $z=0$, and the distortion at $z=2.2$ is dominated by the
geometrical effect.

\begin{figure}
\begin{center}
   \leavevmode\epsfysize=7.5cm \epsfbox{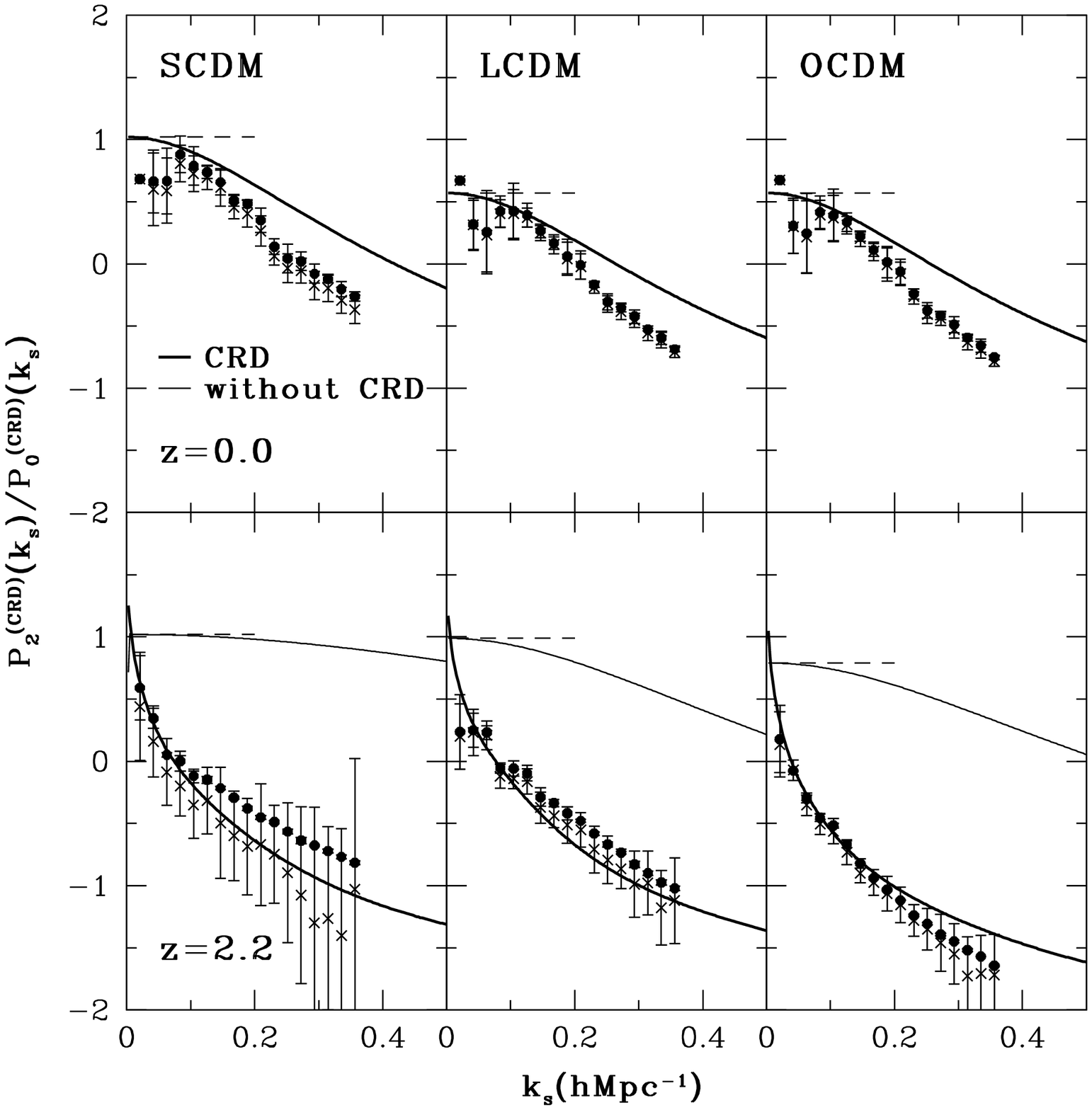}
\end{center}
\vspace*{-1.0cm}
\figcaption{ The quadrupole to monopole ratio of power spectra for
  different cosmologies at $z=0$ (upper panels) and $z=2.2$ (lower
  panels).  Filled circles and crosses correspond to the results from
  N-body simulation, which are evaluated from all particles
  ($1.7\times10^7$) and selected particles($5\times10^4$),
  respectively.  Thick curves correspond to the predictions including
  cosmological redshift distortion effect with an exponential damping
  function. For reference, the predictions neglecting the geometrical
  effect at $z=2.2$ are shown in thin curves.  The dashed lines
  indicate the ratio expected in a linear redshift distortion model.
\label{fig:p2ovp0comb}
}
\end{figure}

Figures \ref{fig:p0combcdf} and \ref{fig:p2combcdf} indicate that the
predictions based on the exponential velocity distribution function
reproduce the simulation results fairly accurately, especially on
large scales (small $k_s$). Also it is interesting and important to
note that even the monopole component is significantly affected by the
cosmological distortion. Figure \ref{fig:p2ovp0comb} displays the
predictions for the quadrupole to monopole ratio. In principle this
statistics is superior either to the monopole or to the quadrupole in
the sense that it is independent of the overall normalization of the
power spectrum (e.g., Cole et al. 1995); in fact the ratio becomes
constant, $(4\beta/3+4\beta^2/7)/(1+2\beta/3+\beta^2/5)$, in linear
regime (i.e., without finger of god and geometrical effects), which is
plotted in dashed lines in the figure. On the other hand, the ratio
lies between $-2 \sim 1$ and the practical usefulness crucially
depends both on the data quality and the accuracy of the theoretical
predictions (see \S 3 below). Incidentally the simulation results seem
to disagree with the predictions at $z=0$, but this is mainly because
the predictions adopt the exponential damping function although the
Gaussian is more appropriate for simulation results at $z=0$ (see also
Figures \ref{fig:p0combcdf} and \ref{fig:p2combcdf}).

\begin{figure}
\begin{center}
   \leavevmode\epsfysize=7.5cm \epsfbox{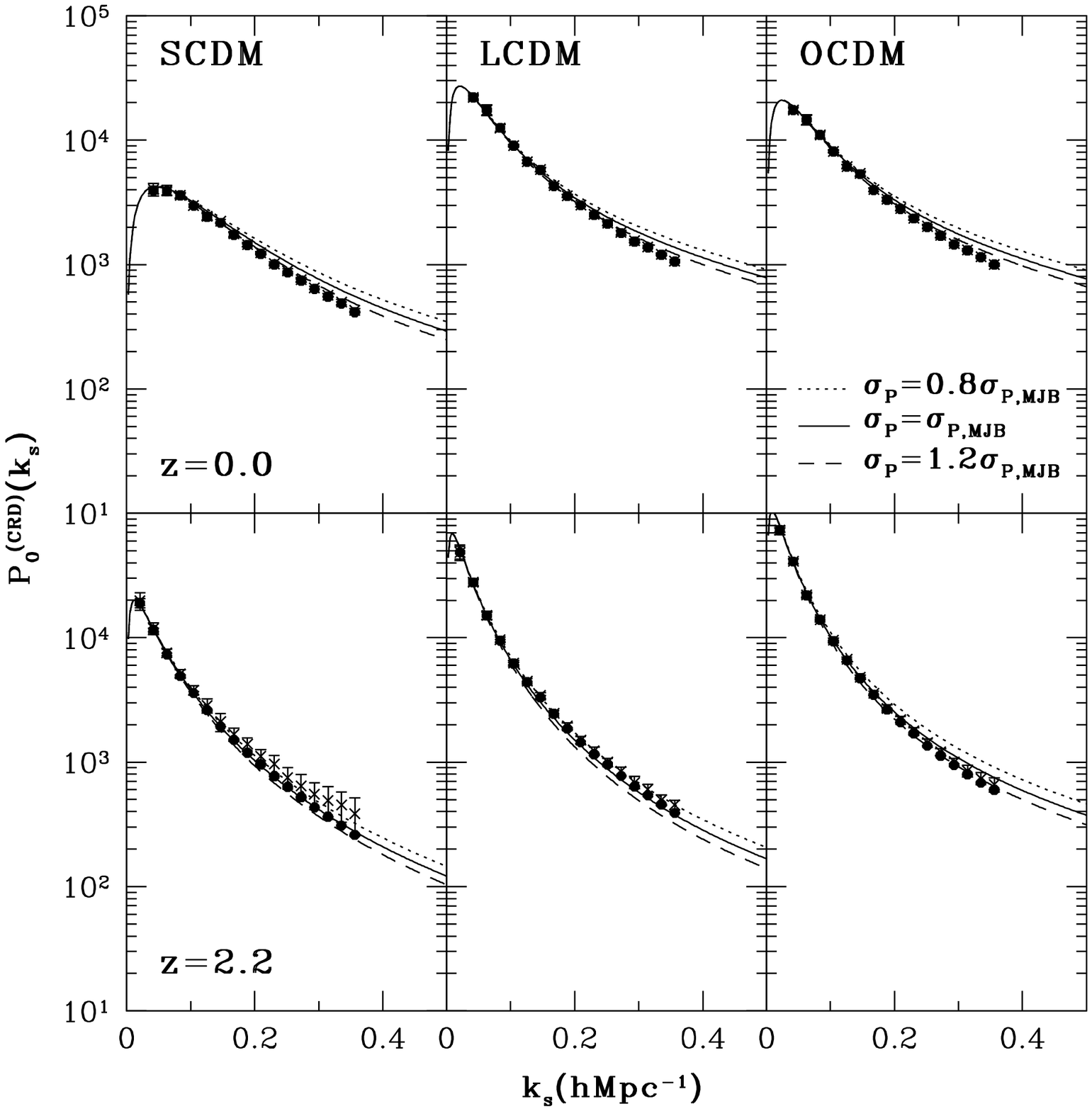}
\end{center}
\vspace*{-1.0cm}
\figcaption{Dependence of the monopole of power spectra on the
  velocity dispersions at $z=0$ (upper panels) and $z=2.2$ (lower
  panels) in CRD space; $\sigmap$ is set to be
  $0.8\sigma_{\scriptscriptstyle{\rm P,MJB}}$ (dotted lines),
  $\sigma_{\scriptscriptstyle{\rm P,MJB}}$ (solid lines), and
  $1.2\sigma_{\scriptscriptstyle{\rm P,MJB}}$ (dashed lines). Filled
  circles and crosses correspond to the results from N-body
  simulation, which are evaluated from all particles ($1.7\times10^7$)
  and selected particles($5\times10^4$), respectively.
 \label{fig:p0combcsp} }
\end{figure}

\begin{figure}
\begin{center}
   \leavevmode\epsfysize=7.5cm \epsfbox{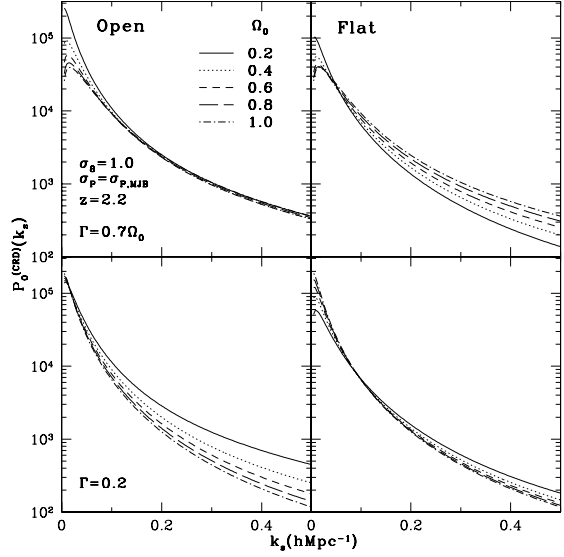}
\end{center}
\figcaption{Dependence of the monopole of power spectra in
 cosmological redshift space at $z=2.2$ on $\Omega_0$.  We consider both
 open ($\lambda_0=0$) and spatially-flat ($\lambda_0=1-\Omega_0$) models
 in left and right panels, respectively.  The shape parameter $\Gamma$
 is fixed as $0.7\Omega_0$ in upper, and $0.2$ in lower panels.
 The fluctuation amplitude $\sigma_8$ is fixed as unity for simplicity.
 \label{fig:p0cosmo} }
\end{figure}

\begin{figure}
\begin{center}
   \leavevmode\epsfysize=7.5cm \epsfbox{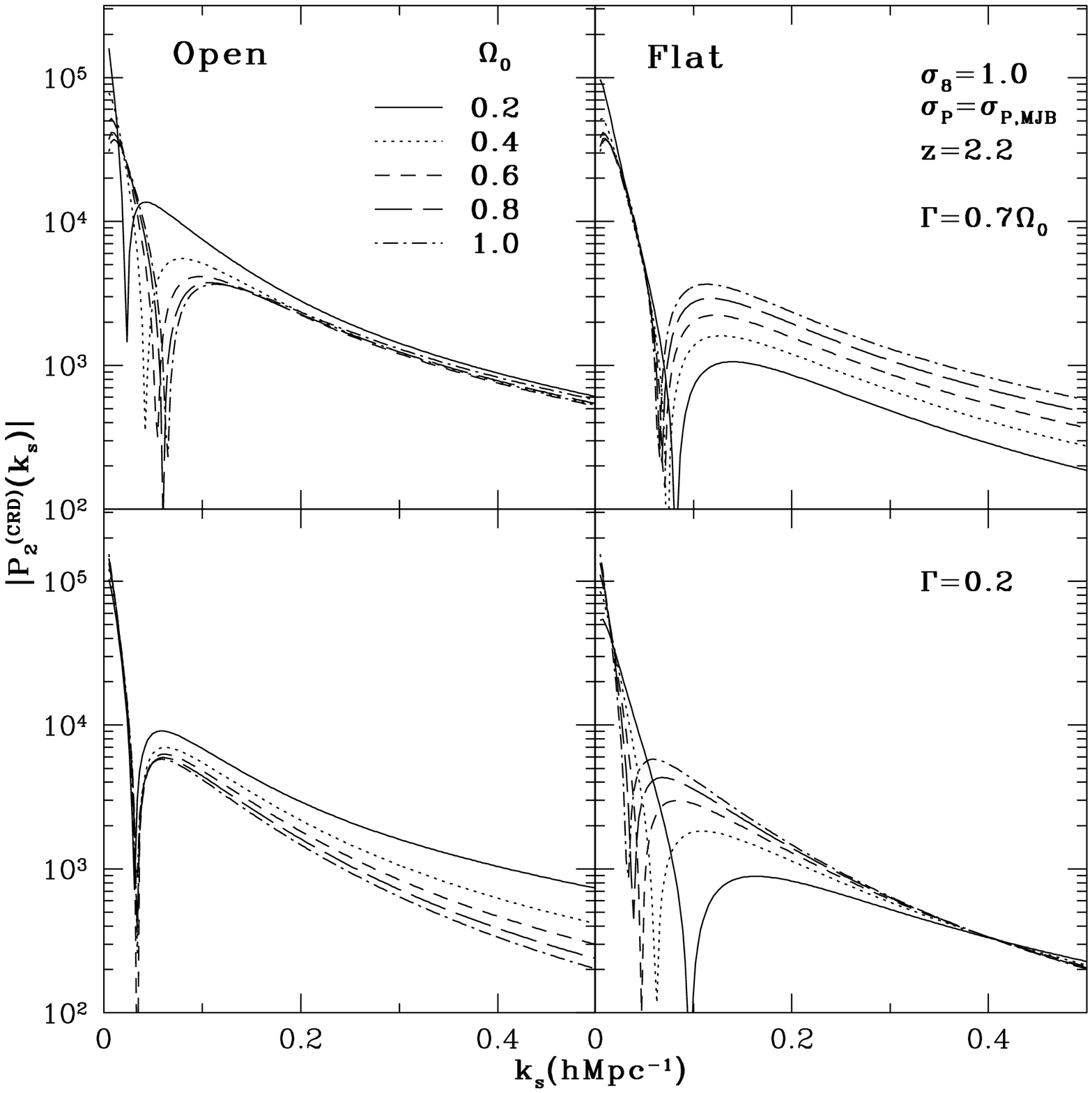}
\end{center}
\figcaption{The same as Figure \protect\ref{fig:p0cosmo} but for
the quadrupole moment of the power spectra.
\label{fig:p2cosmo}
}
\end{figure}

\begin{figure}
\begin{center}
   \leavevmode\epsfysize=7.5cm \epsfbox{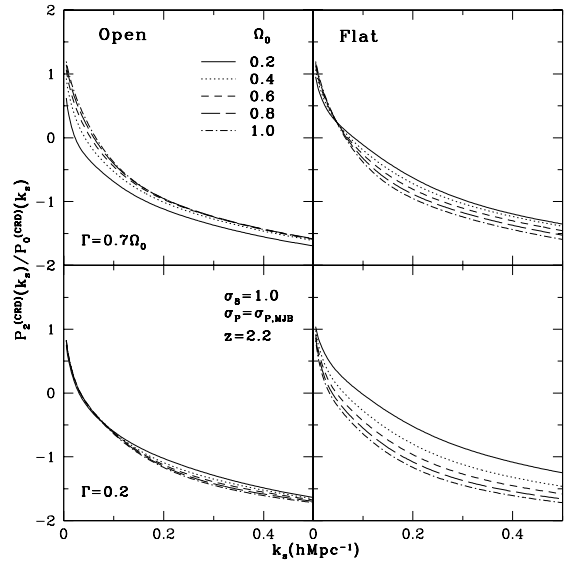}
\end{center}
\figcaption{ The same as Figure \protect\ref{fig:p0cosmo} but for the
  quadrupole to monopole ratio.
\label{fig:p2ovp0cosmo}
}
\end{figure}

In order to understand the model-dependence of the moments, we show
the extent to which it is sensitive to the assumed velocity
dispersions $\sigmap$ in Figure \ref{fig:p0combcsp} for
$P^{(\CRD)}_0(k_s)$, and to a set of parameters ($\Omega_0$,
$\lambda_0$, and $\Gamma$) in Figures \ref{fig:p0cosmo},
\ref{fig:p2cosmo} and \ref{fig:p2ovp0cosmo} for $P^{(\CRD)}_0(k_s)$,
$P^{(\CRD)}_2(k_s)$, and $P^{(\CRD)}_2(k_s)/P^{(\CRD)}_0(k_s)$,
respectively.  Figure \ref{fig:p0combcsp} implies that the predictions
on scales $k_s\simlt 0.2h {\rm Mpc}^{-1}$ are very accurate and not
sensitive to the adopted $\sigmap$. Thus the usefulness of our
proposed cosmological test is crucially dependent on the observational
data quality on the quasi-linear scales which are presumably
achievable only with the upcoming SDSS QSO surveys.  It should be
noted that there is a small but clear systematic difference between
$N=1.7\times10^7$ and $5\times10^4$ in Figures \ref{fig:p0combcdf} to
\ref{fig:p0combcsp}, despite the fact that we subtracted the shot
noise due to the finite number of sampled particles. Actually we were
not able to understand the origin of this effect, but the effect is
smaller than the other uncertainties involved in the present analysis.

Incidentally Figure \ref{fig:p0cosmo} indicates that the behavior of
$P^{(\CRD)}_0(k_s)$ is degenerate for a certain set of the parameters
since its shape and amplitude are sensitive both to $\Omega_0$ and
$\lambda_0$ through the correction factors, $c_\perp(z)$ and
$c_\parallel(z)$, and to $\Gamma$ through the shape of the power
spectrum in real space. Such a degeneracy on the density parameter is
apparent for open models of $\Gamma=0.7\Omega_0$ and for flat models
of $\Gamma=0.2$. Fortunately, as shown by Figure \ref{fig:p2cosmo}, a
measurement of the quadrupole $P^{(\CRD)}_2(k_s)$ may help break the
degeneracy especially for the density parameter as low as $\sim
0.2$. Combining other cosmological tests, such as the cosmic microwave
background, clustering statistics, cluster abundances, etc. may
further narrow the space of the cosmological parameters. In the next
section we will investigate the feasibility of measuring the
cosmological parameters by combining the first two moments of the
red-shift power spectrum from a survey like the Sloan QSO survey and the
cluster abundance.

\section{Feasibility of determining the cosmological parameters
\protect \\ \protect
from the power spectrum in cosmological redshift space}
\label{sec:feasibility}

In the previous section we have demonstrated that the theoretical
predictions for $P^{(\CRD)}_0(k_s;z)$ and $P^{(\CRD)}_2(k_s;z)$ are quite
accurate for $k_s\simlt0.2h$Mpc$^{-1}$. Thus we finally examine if the
cosmological test using these moments leads to any useful constraints on
the cosmological parameters.  For this purpose, we use the results from
the N-body simulation models (Table \ref{tab:modelparam}) at $z=2.2$,
and perform the statistical analyses as follows.

Our theoretical model is specified by a set of $\Omega_0$,
$\lambda_0$, $\sigma_8$, and $h$. Since the results of our N-body
simulations are scalable with respect to $h$, we assume that $h$ is related
to the shape of the spectrum through the relation, $h \equiv
\Gamma/\Omega_0$, neglecting the baryon contribution $\Omega_b$
(Sugiyama 1995) for simplicity. In order to take account of the
statistical limitation, we do not use the entire simulation particles,
but randomly select particles. For a given number of selected
particles, we generate 24 mock samples (8 randomly selected subsamples
for each realization) for each cosmological model. Then we assign the
errors to the simulation data from the ensemble average over the 24
samples. In order to avoid the strongly nonlinear effect, we use the
spectrum in the range $k_s\simlt0.2h$Mpc$^{-1}$, and perform the
$\chi^2$-analysis.

\begin{figure}
\begin{center}
   \leavevmode\epsfysize=7.5cm \epsfbox{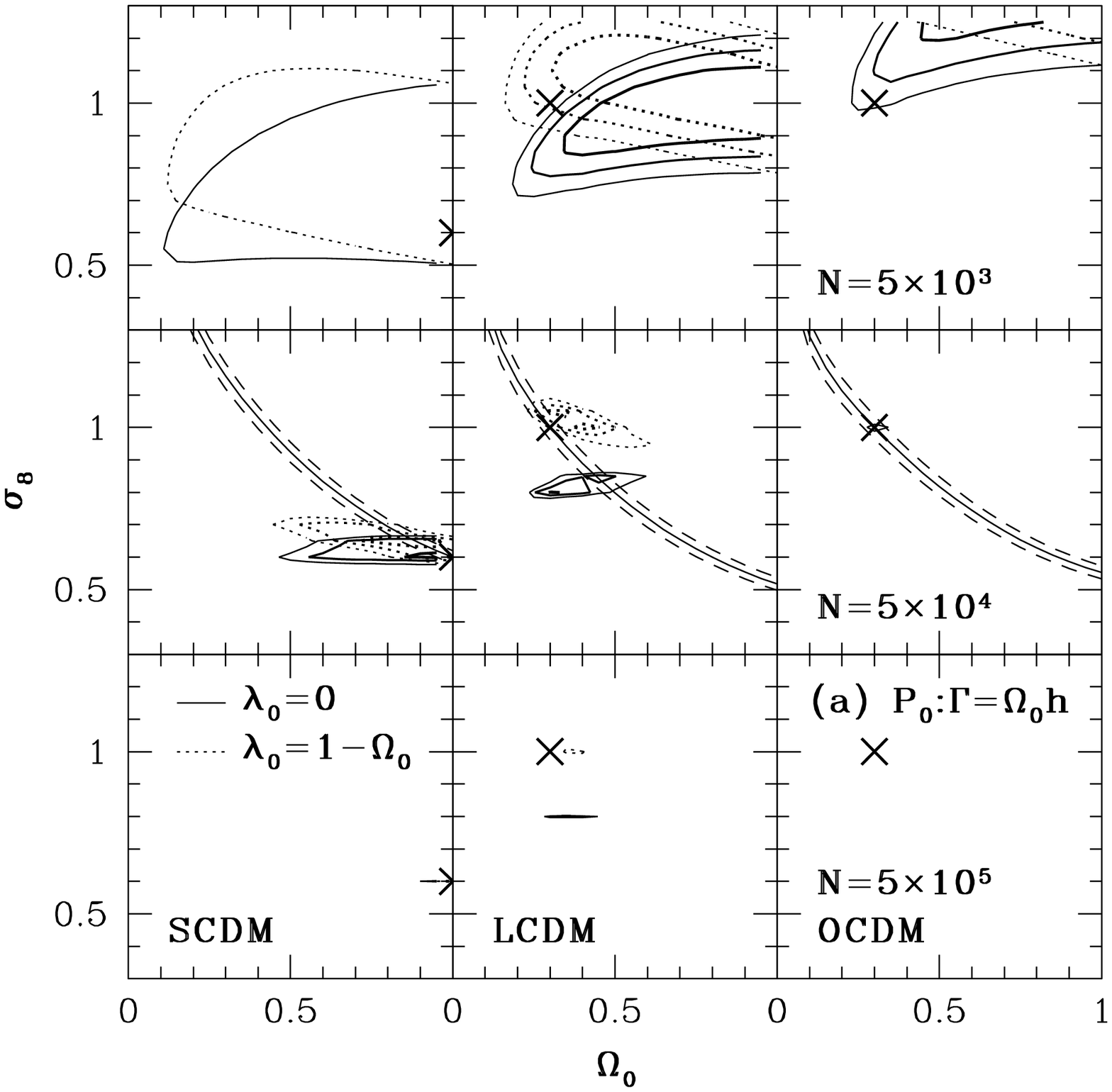}
   \leavevmode\epsfysize=7.5cm \epsfbox{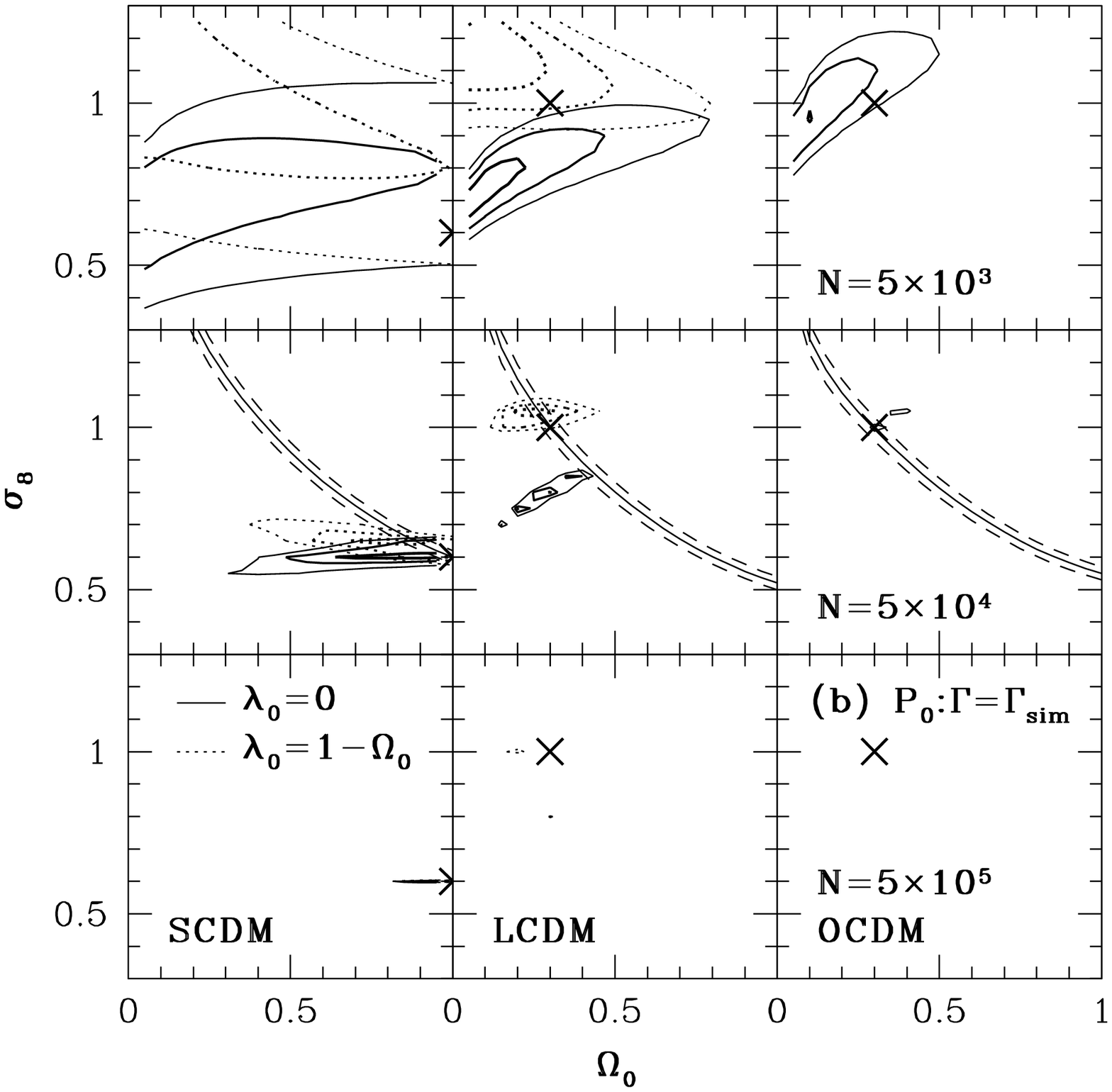}
\vspace*{0.5cm}
   \leavevmode\epsfysize=7.5cm \epsfbox{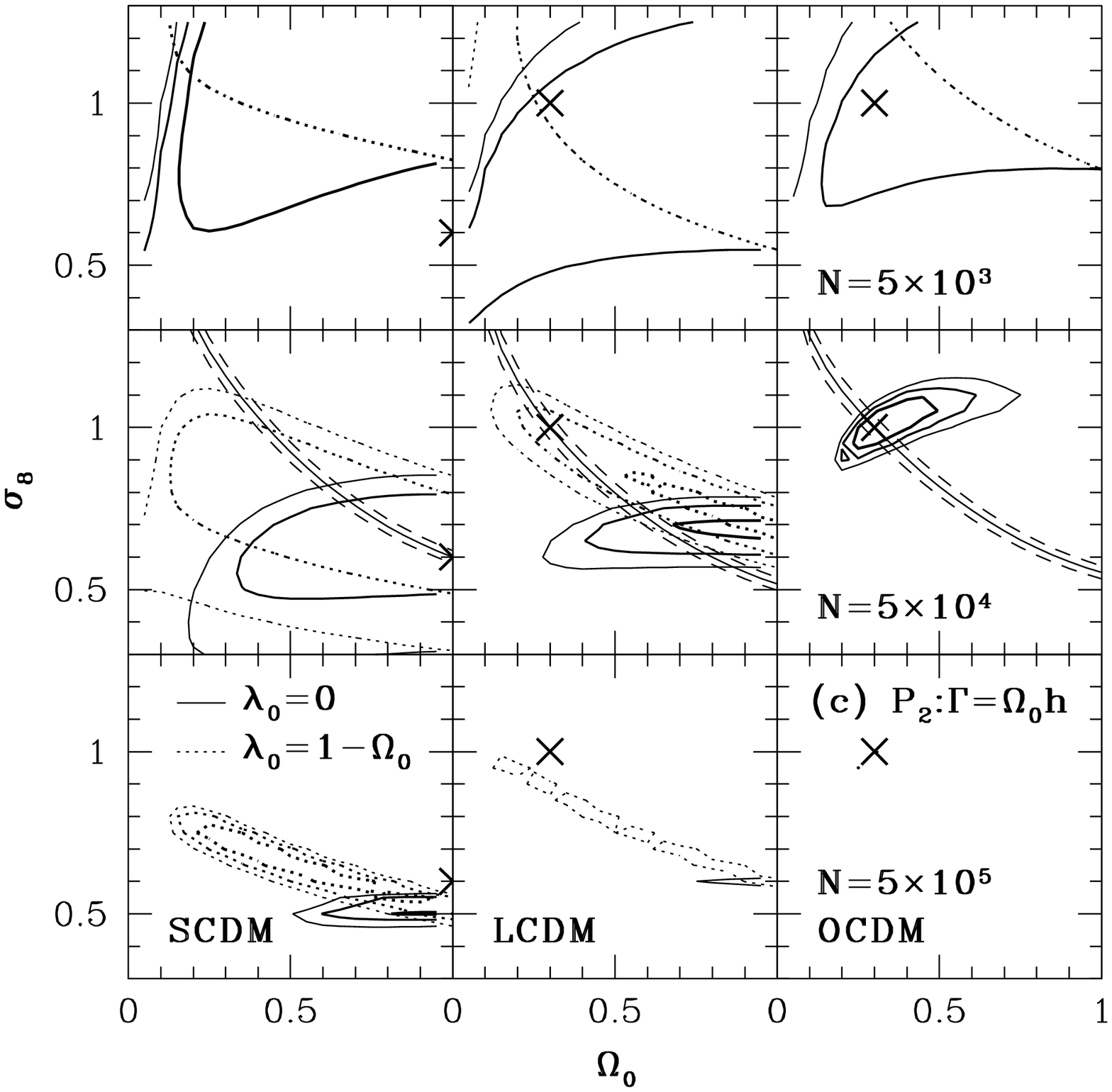}
   \leavevmode\epsfysize=7.5cm \epsfbox{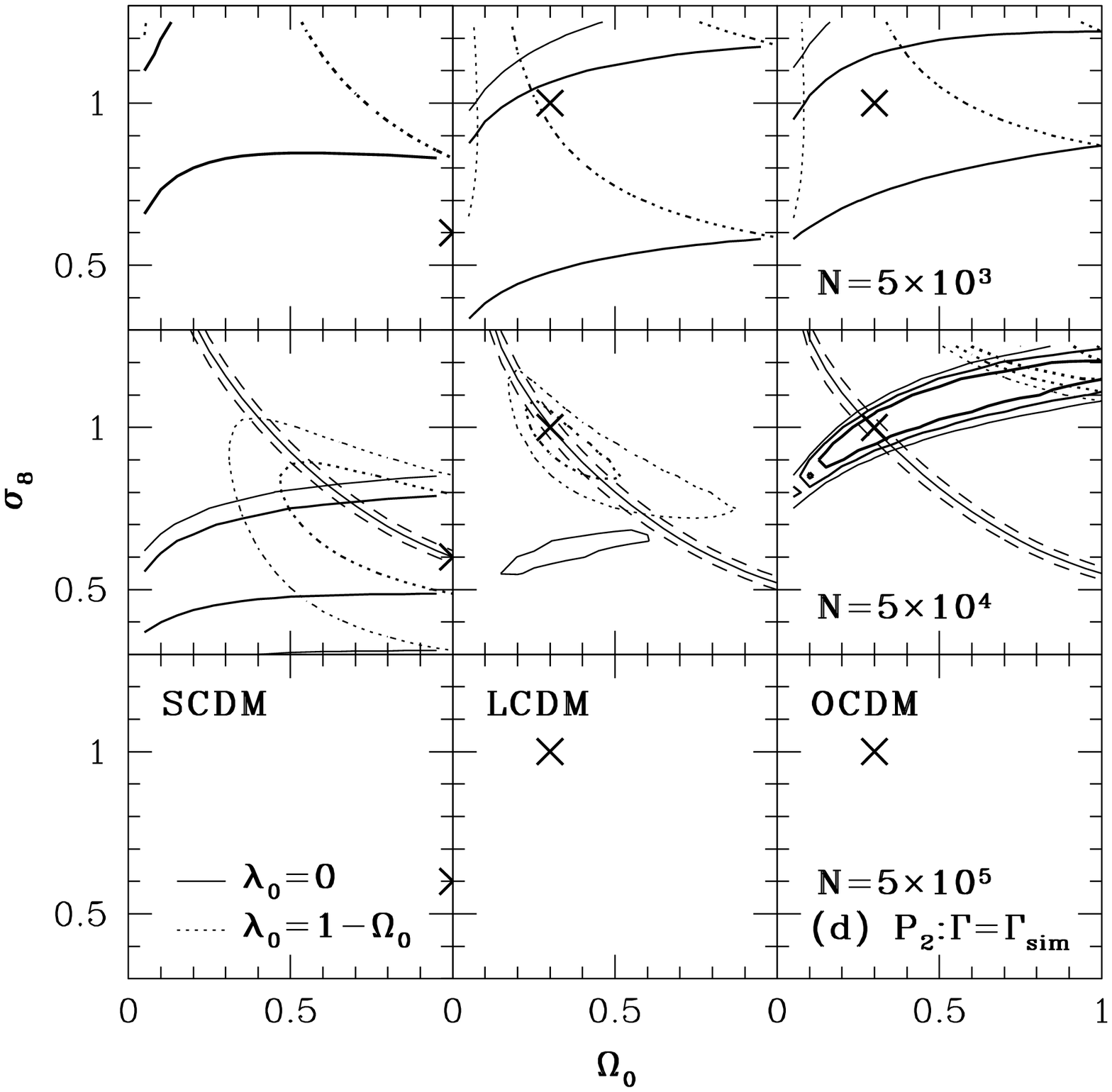}
\end{center}
\figcaption{The confidence contours on $\Omega_0$-$\sigma_8$ plane
  from the $\chi^2$-analysis of the monopole and quadrupole moments of
  the power spectrum in the cosmological redshift space at $z=2.2$;
  (a) $\Gamma=\Omega_0 h$ for $P_0^{(\CRD)}(k_s)$, (b)
  $\Gamma=\Gamma_{sim}$ for $P_0^{(\CRD)}(k_s)$, (c) $\Gamma=\Omega_0
  h$ for $P_2^{(\CRD)}(k_s)$, (d) $\Gamma=\Gamma_{sim}$ for
  $P_2^{(\CRD)}(k_s)$.  The long-dashed, heavy thick, light thick, and
  thin lines correspond to 0.5, 1, 2, and 3$\sigma$ confidence levels.
  We randomly selected $N=5\times10^3$ (upper panels), $N=5\times10^4$
  (middle panels), and $N=5\times10^5$ (lower panels) particles from
  N-body simulation. Solid and dotted lines represent results for open
  and spatially-flat models, respectively.  The crosses indicate the
  true values of our simulations.  We also show the constraint on
  $\Omega_0$ and $\sigma_8$ from the cluster abundance by thin lines
  in the middle panels.
  \label{fig:chi2p0p2os8} }
\end{figure}

\begin{figure}
\begin{center}
   \leavevmode\epsfysize=7.5cm \epsfbox{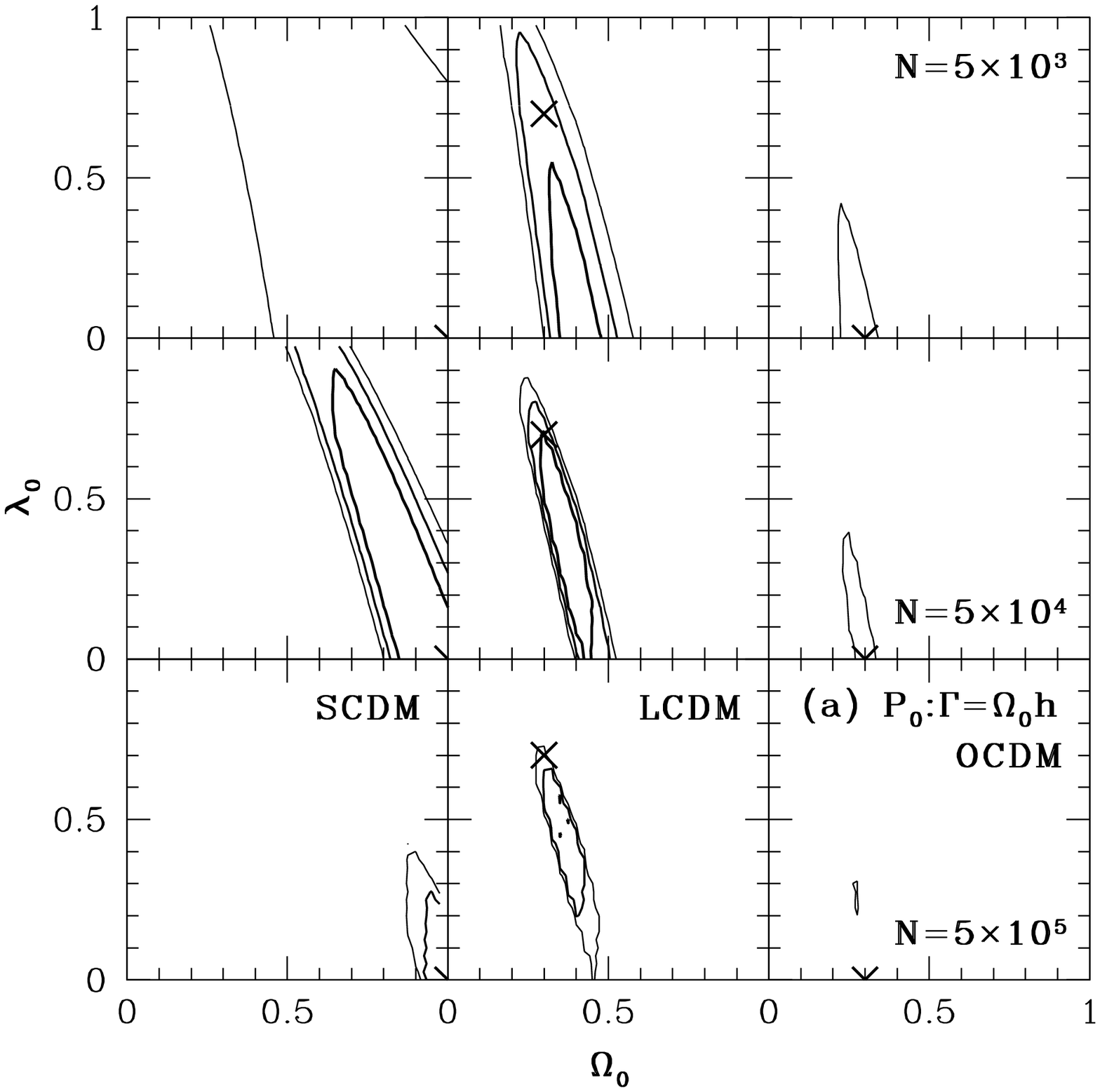}
   \leavevmode\epsfysize=7.5cm \epsfbox{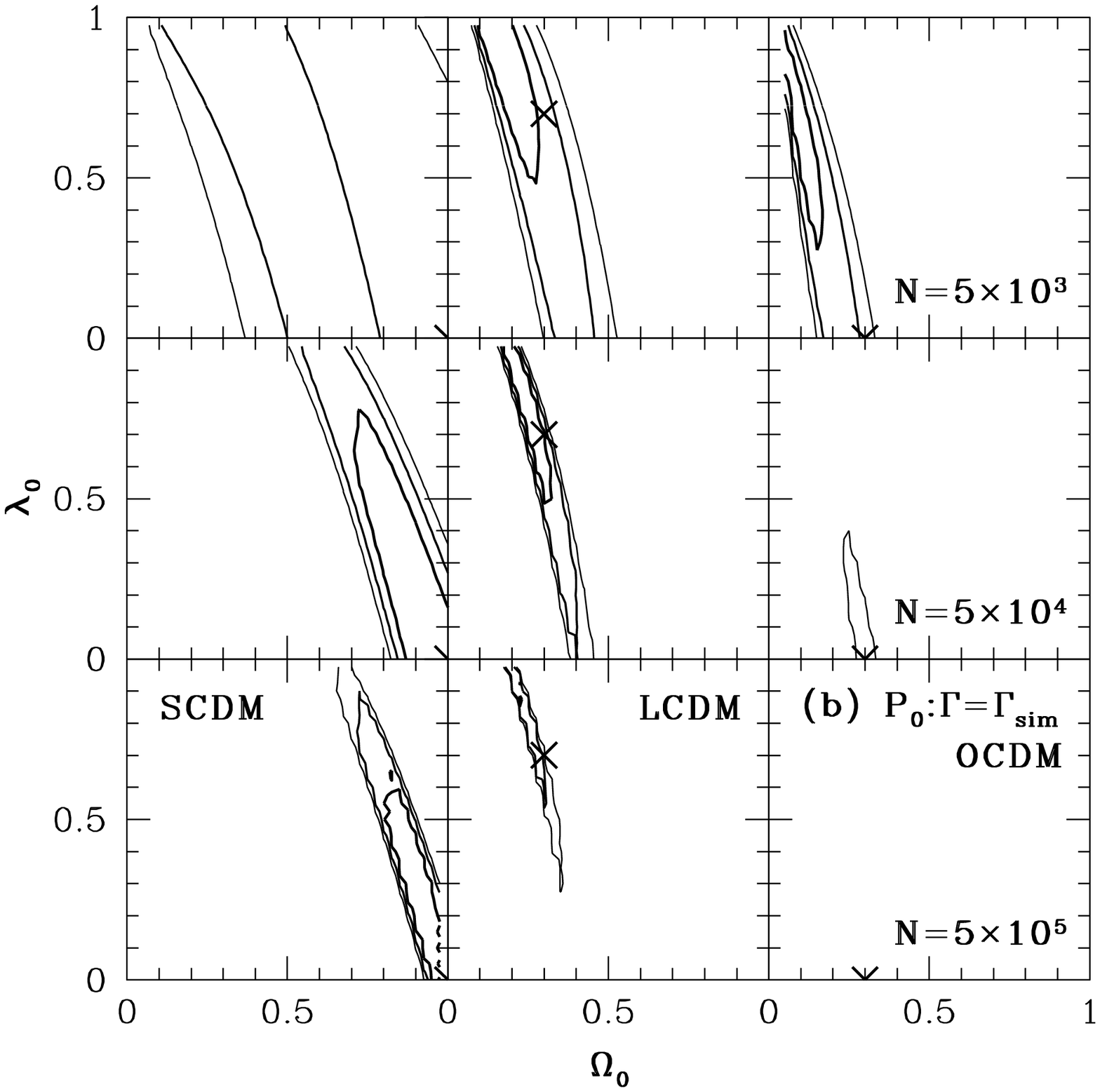}
\vspace*{0.5cm}
   \leavevmode\epsfysize=7.5cm \epsfbox{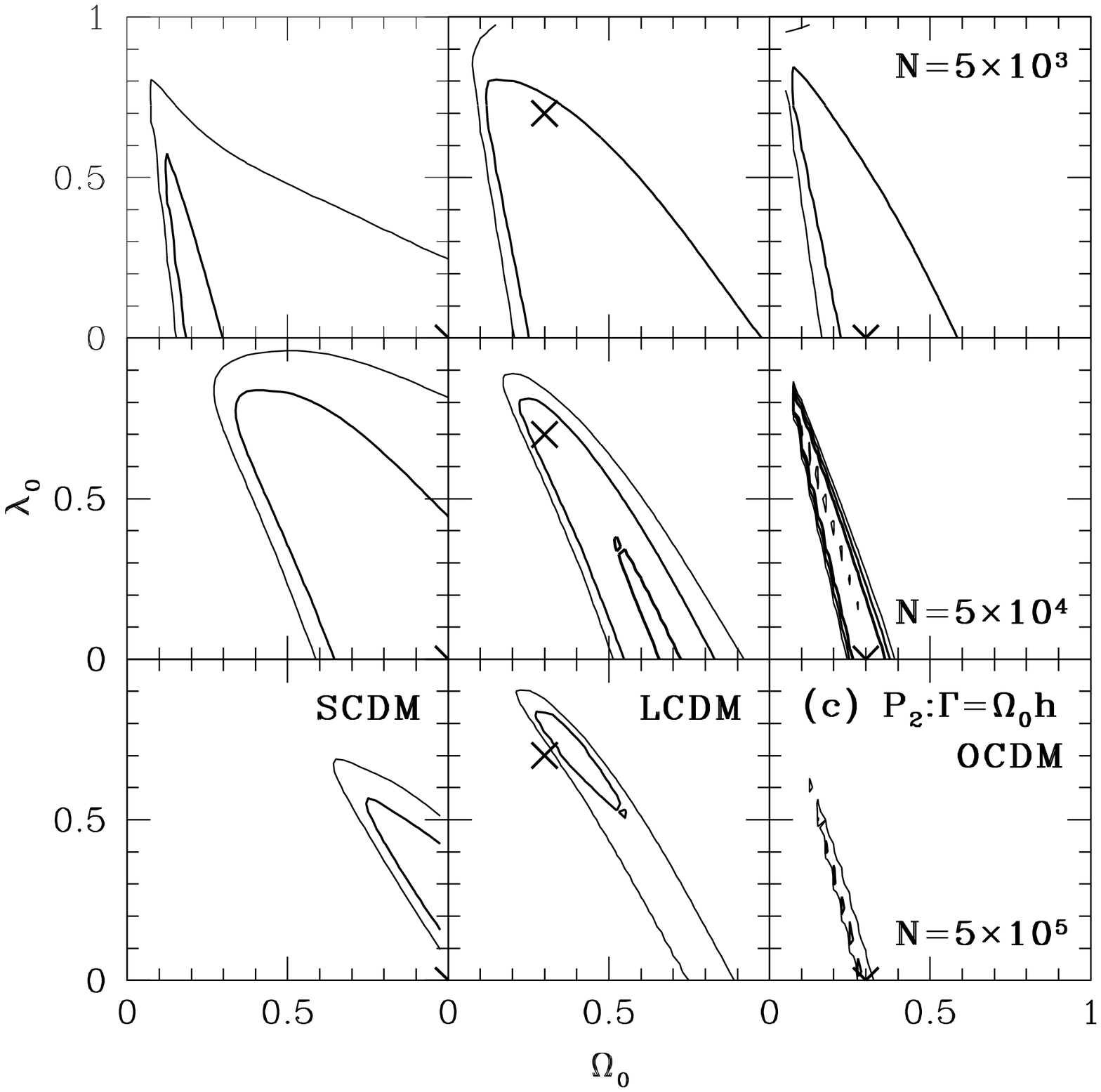}
   \leavevmode\epsfysize=7.5cm \epsfbox{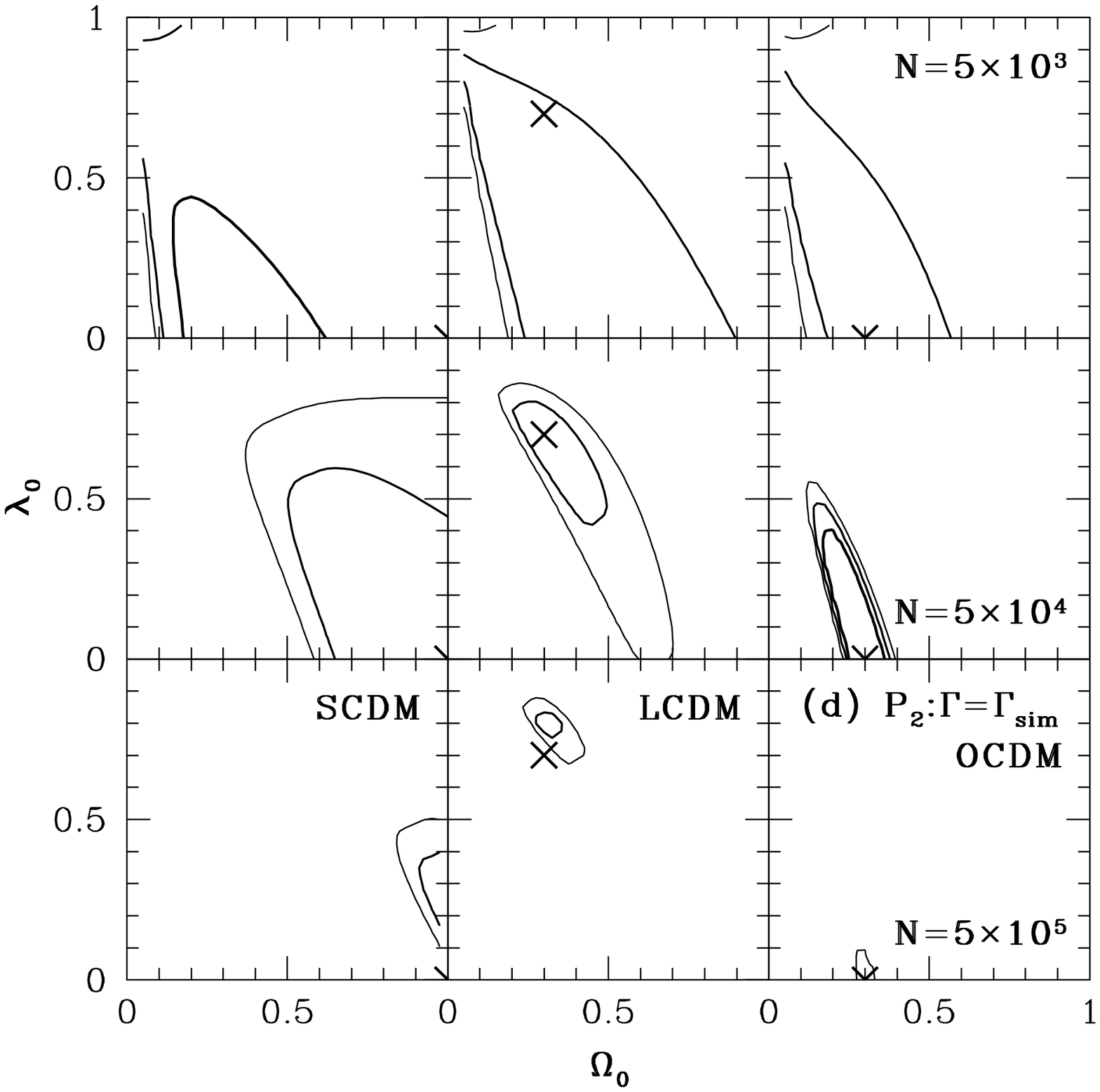}
\end{center}
\figcaption{The same as Figure \protect\ref{fig:chi2p0p2os8} but on
  $\Omega_0$-$\lambda_0$ plane.  We adopt the value of $\sigma_8$ from
  equation (\protect\ref{eq:s8o0}\protect) with the quoted error bars.
\label{fig:chi2p0p2ol}
}
\end{figure}

Figures \ref{fig:chi2p0p2os8} and \ref{fig:chi2p0p2ol} display the
confidence level contours from the $\chi^2$-analysis of
$P^{(\CRD)}_0(k_s;z)$ or $P^{(\CRD)}_2(k_s;z)$ on $\Omega_0$--$\sigma_8$
and $\Omega_0$--$\lambda_0$ planes, respectively.  To be specific, we
compute the $\chi^2$ defined by
\begin{equation}
\label{eq:chi2def}
  \chi^2 =\sum_{i=1}^{10}
\left[\frac{P_{\rm sim}^{R,S}(k_{s,i})-P_{\rm model}(k_{s,i})}
  {\Delta_{\P {\rm sim}}(k_{s,i})}\right]^2 ,
\end{equation}
where the wavenumber is sampled between $0.02h$Mpc$^{-1}\simlt k_s
\simlt 0.2h$Mpc$^{-1}$ as
\begin{equation}
  k_{s,i}=\frac{2\pi}{L_{\rm box}}i \quad (i=1,\cdots,10) .
\end{equation}
The indices $R$ and $S$ denote the three different realizations and
eight randomly selected subsamples for each cosmological model, and in
evaluating equation (\ref{eq:chi2def}) we randomly choose one mock
sample from 24 ($=3\times 8$) simulated mock samples in total.  For
$\Delta^2_{\P {\rm sim}}(k_{s,i})$, we use the dispersion among 24
mock samples assuming that this corresponds to a cosmic variance:
\begin{eqnarray}
\bar P_{\rm sim}(k_{s,i}) &=& \frac{1}{24}
\sum_{R=1}^{3} \sum_{S=1}^{8} P_{\rm sim}^{R,S}(k_{s,i})\\
\Delta^2_{\P {\rm sim}}(k_{s,i}) &=& \frac{1}{24-1}
\sum_{R=1}^{3} \sum_{S=1}^{8} 
[P_{\rm sim}^{R,S}(k_{s,i})-\bar P_{\rm sim}(k_{s,i})]^2 .
\end{eqnarray}
The resulting confidence level is computed from the reduced $\chi^2$
for eight degrees of freedom, i.e., ten data points {\it minus} two
free parameters, either, $(\Omega_0,\sigma_8)$ or
$(\Omega_0,\lambda_0)$.  We perform the above analysis for the
monopole and the quadrupole, separately, and also combine the monopole
and quadrupole analysis.  We have adopted the $\chi^2$ technique to
quantify the errors for the measured parameters because it is simple,
but perhaps other statistical methods, e.g., the maximum likelihood,
may better quantify the measurement errors in the actual application
to the future observational data.

Figures \ref{fig:chi2p0p2os8}a, \ref{fig:chi2p0p2os8}c,
\ref{fig:chi2p0p2ol}a, and \ref{fig:chi2p0p2ol}c assume that
$\Gamma=\Omega_0 h$, while other panels in these two figures treat
$\Gamma$ as an independent parameter and fix the value as specified by
the simulation (Table \ref{tab:modelparam}).  We display the results
for $N=5\times10^5$ (bottom panels), $5\times10^4$ (middle panels),
and $5\times10^3$ (top panels) from the entire $256^3$ particles at
$z=2.2$. For reference, the quasar luminosity function of Boyle,
Shanks, \& Peterson (1988) predicts that the number of quasars per
$\pi$ steradian brighter than 19 magnitude in B is about 4500 either
in $z=0.9\sim 1.1$ or in $z=1.9\sim 2.1$ (for the $\Omega_0=1$ and
$\lambda_0=0$ model). If we use the extrapolation of the luminosity
function to $z\sim 5$ by Wallington \& Narayan (1993), the total
number of $\sim 10^5$ QSOs $0\simlt z \simlt 5$ are expected to be
cataloged in the upcoming SDSS QSO sample.  It should be noted that
the best-fit parameters in those figures are sometimes a bit different
from the true values that we use in the simulations. This is simply
because the figures represent results for one particular mock sample.
We made sure that the best-fit values are in good agreement with the
true values if we replace $P_{\rm sim}^{R,S}(k_{s,i})$ in equation
(\ref{eq:chi2def}) by $\bar P_{\rm sim}(k_{s,i})$.

\begin{figure}
\begin{center}
   \leavevmode\epsfysize=7.5cm \epsfbox{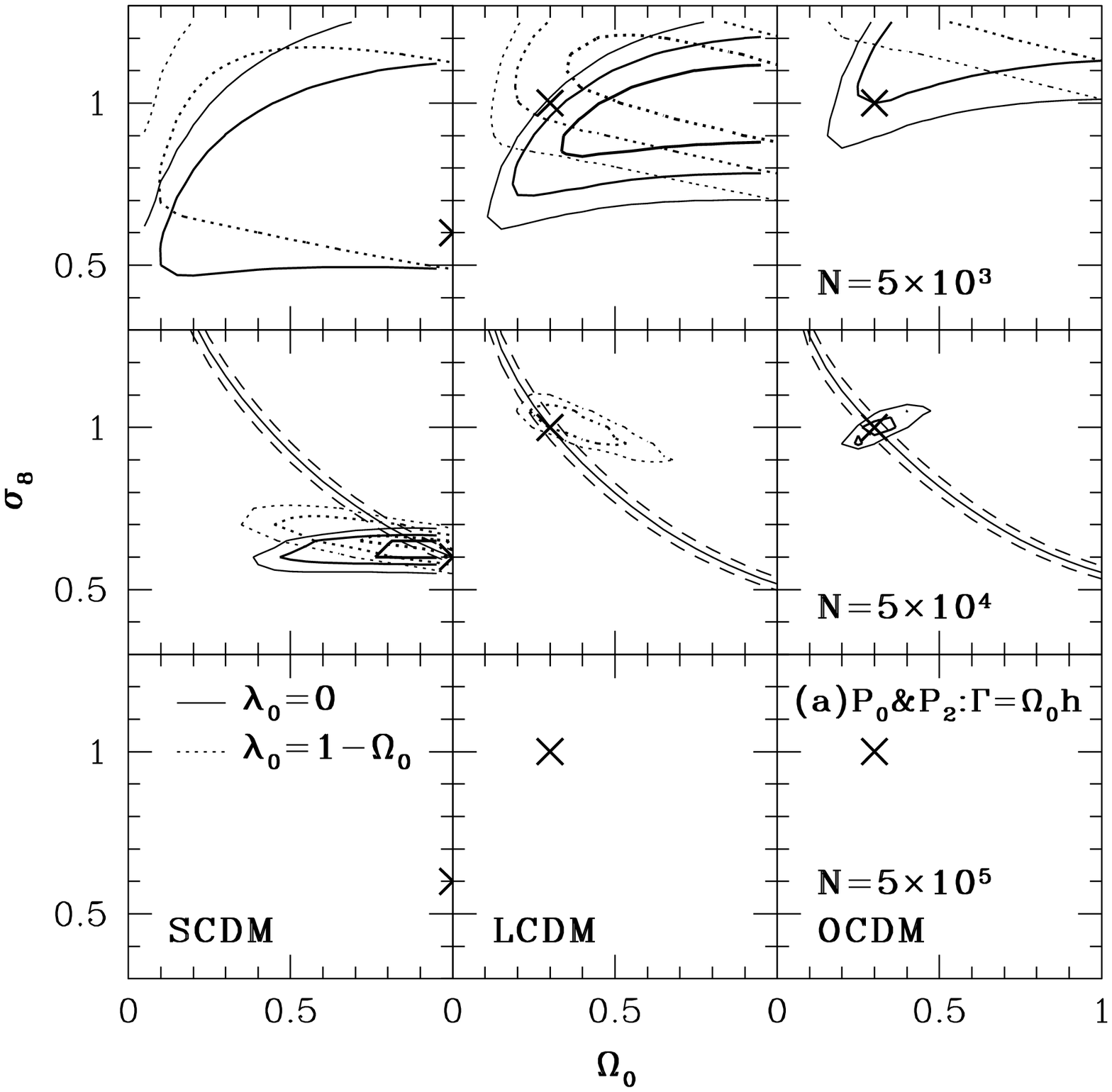}
\vspace*{0.5cm}
   \leavevmode\epsfysize=7.5cm \epsfbox{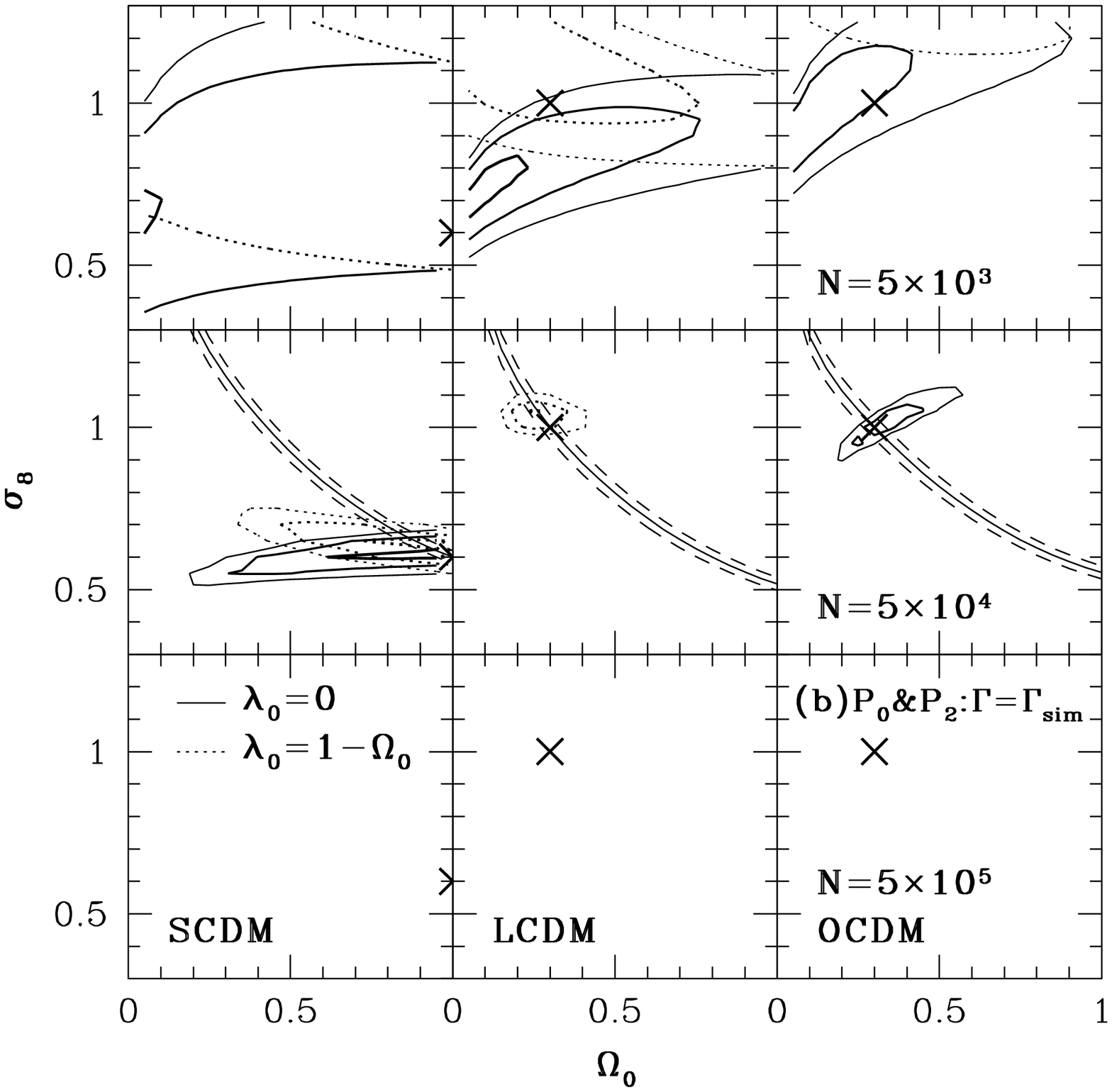}
\end{center}
\figcaption{
The same as Figure \protect\ref{fig:chi2p0p2os8} but from
  the combined analysis of the monopole and quadrupole moments.
  \label{fig:chi2p02os8} }
\end{figure}

We repeat the analysis in Figure \ref{fig:chi2p0p2os8} by fixing
either $\lambda_0=0$ or $=1-\Omega_0$. As anticipated, the present
analyses with either $P^{(\CRD)}_0(k_s)$ or $P^{(\CRD)}_2(k_s)$ alone do
not determine the cosmological parameters completely.  However, the
error contours of the monopole do not align with those of the
quadrupole completely; in some cases the contours of the monopole and the
quadrupole are somewhat orthogonal, though the error contours of the
quadrupole are generally larger than those of the monopole. These
results indicate that a combination of the first two moments can
more stringently, as expected, constrain the cosmological
parameters. Figures \ref{fig:chi2p02os8} show the confidence levels of
such a joint constraint on $(\sigma_8,\Omega_0)$.  An encouraging
feature noted from this figure is that the flat and open universes
with $\Omega_0=0.3$ can be well separated with this test if the number
of the objects $N\ge 5\times 10^{4}$. Nevertheless the combination
with another cosmological test can easily yield an interesting
constraint in the parameter space. This is illustrated in the middle
panels of Figures \ref{fig:chi2p0p2os8} and \ref{fig:chi2p02os8} where
the constraints from the X-ray cluster abundance (Kitayama \& Suto
1997):
\begin{eqnarray}
  \sigma_8 = (A \pm 0.02 ) \times \left\{
      \begin{array}{ll}
        \Omega_0^{-0.35-0.82\Omega_0+0.55\Omega_0^2} &
        \mbox{($\lambda_0=1-\Omega_0$)}, \\ 
        \Omega_0^{-0.28-0.91\Omega_0+0.68\Omega_0^2} &
        \mbox{($\lambda_0=0$)} . 
      \end{array}
   \right. 
\label{eq:s8o0}
\end{eqnarray}
are overlaid. The normalization factor $A$ determined from the
observed cluster abundance is 0.54. 

\begin{figure}
\begin{center}
   \leavevmode\epsfysize=7.5cm \epsfbox{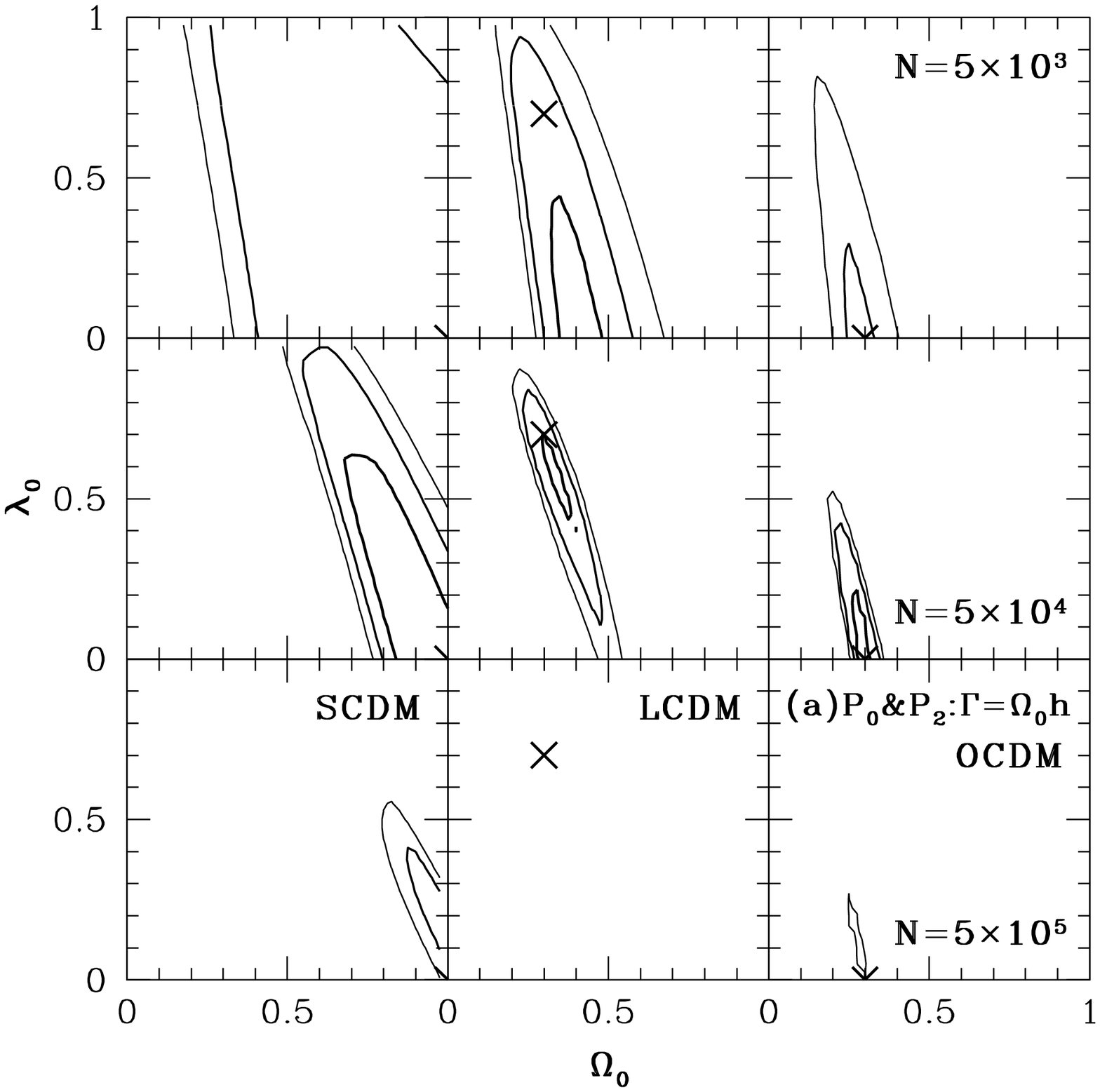}
   \leavevmode\epsfysize=7.5cm \epsfbox{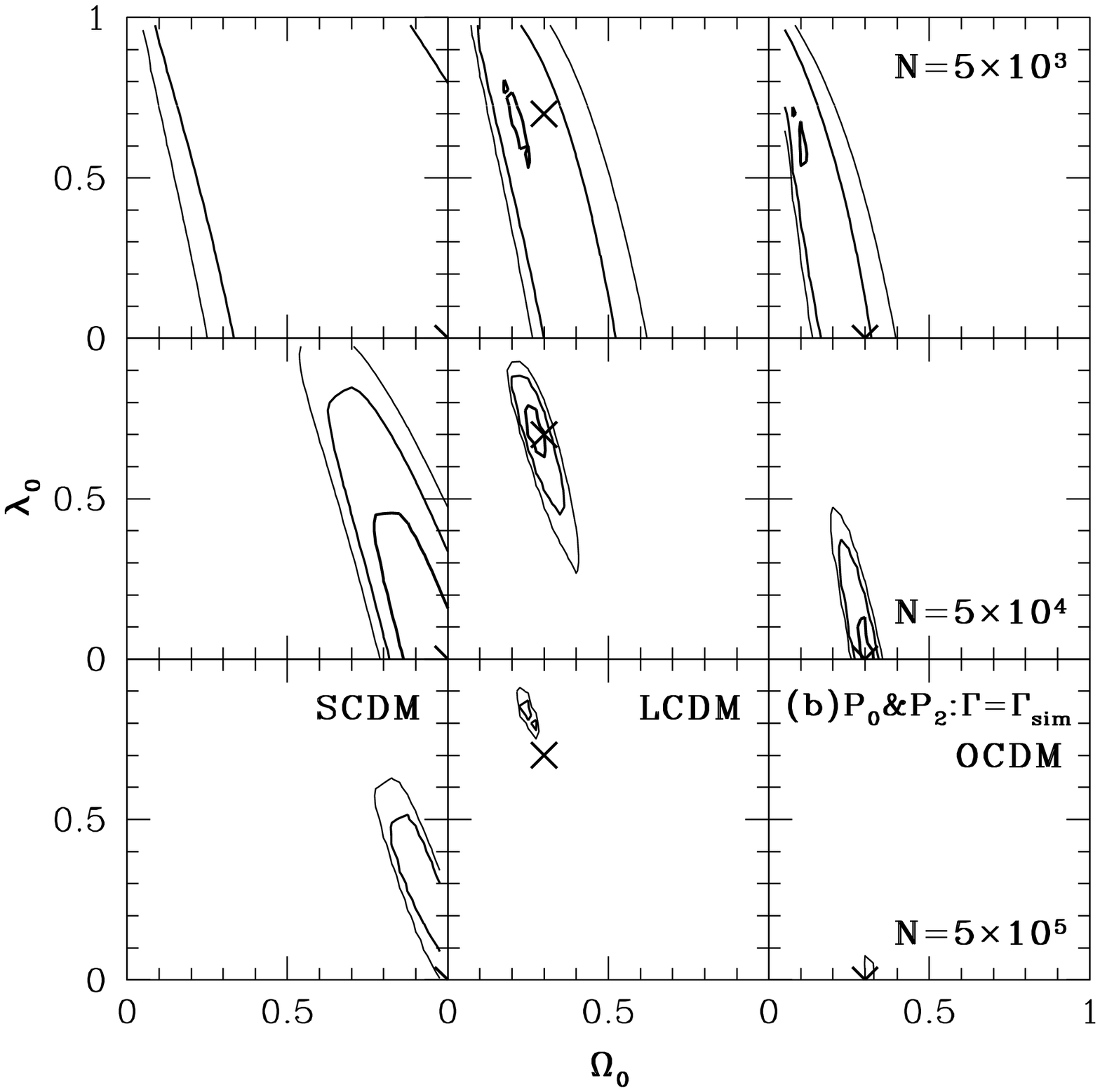}
\end{center}
\figcaption{
The same as Figure \protect\ref{fig:chi2p0p2ol} but from
  the combined analysis of the monopole and quadrupole moments.
\label{fig:chi2p02ol}
}
\end{figure}

Combining the cluster abundance with the test of this paper may
determine the cosmological parameters $\lambda_0$ and $\Omega_0$. This
is illustrated in Figure \ref{fig:chi2p0p2ol}.  The normalization
$\sigma_8$ adopted in our simulations are slightly different from the
value given by Eq.(\ref{eq:s8o0}). Thus in the analysis presented in
Figures \ref{fig:chi2p0p2ol} and \ref{fig:chi2p02ol}, we accordingly
change the normalization factor $A$ with keeping the same $\Omega_0$
dependence, and find the minimal $\chi^2$ by allowing the $\pm 0.02$
dispersion.  Our adopted value of $A$ for each model is summarized in
Table \ref{tab:modelparam}. The monopole moment is sensitive to the
density parameter $\Omega_0$ only, while the quadrupole measurement
interestingly complements to the monopole measurement in that it
depends more strongly on the cosmological constant. Combining these
two moments results in a joint constraint on $(\Omega_0, \lambda_0)$
which is shown in \ref{fig:chi2p02ol}. Consistent with the results
shown in Figure \ref{fig:chi2p02os8}, the open and flat universes with
$\Omega_0 \simlt 0.3$ can be discriminated with the number of objects
larger than $5\times 10^{4}$.  It is also interesting to note that our
constraints in Figure \ref{fig:chi2p02ol} are fairly orthogonal to
those from the SN Ia (Perlmutter et al.  1999), and can be even
combined to probe the cosmic equation of state in general (e.g.,
Garnavich et al. 1998). 

\begin{figure}
\begin{center}
   \leavevmode\epsfysize=7.5cm \epsfbox{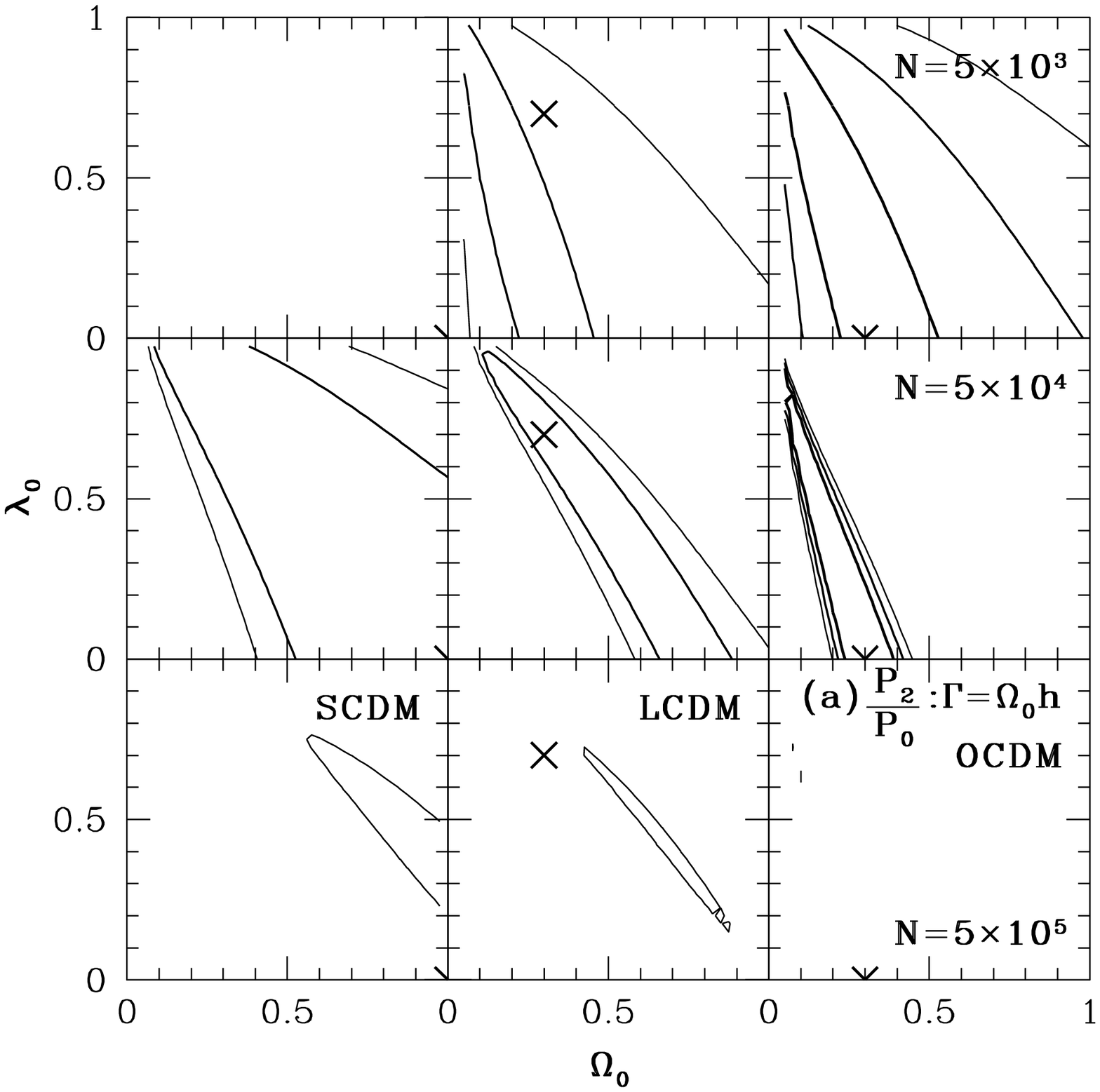}
   \leavevmode\epsfysize=7.5cm \epsfbox{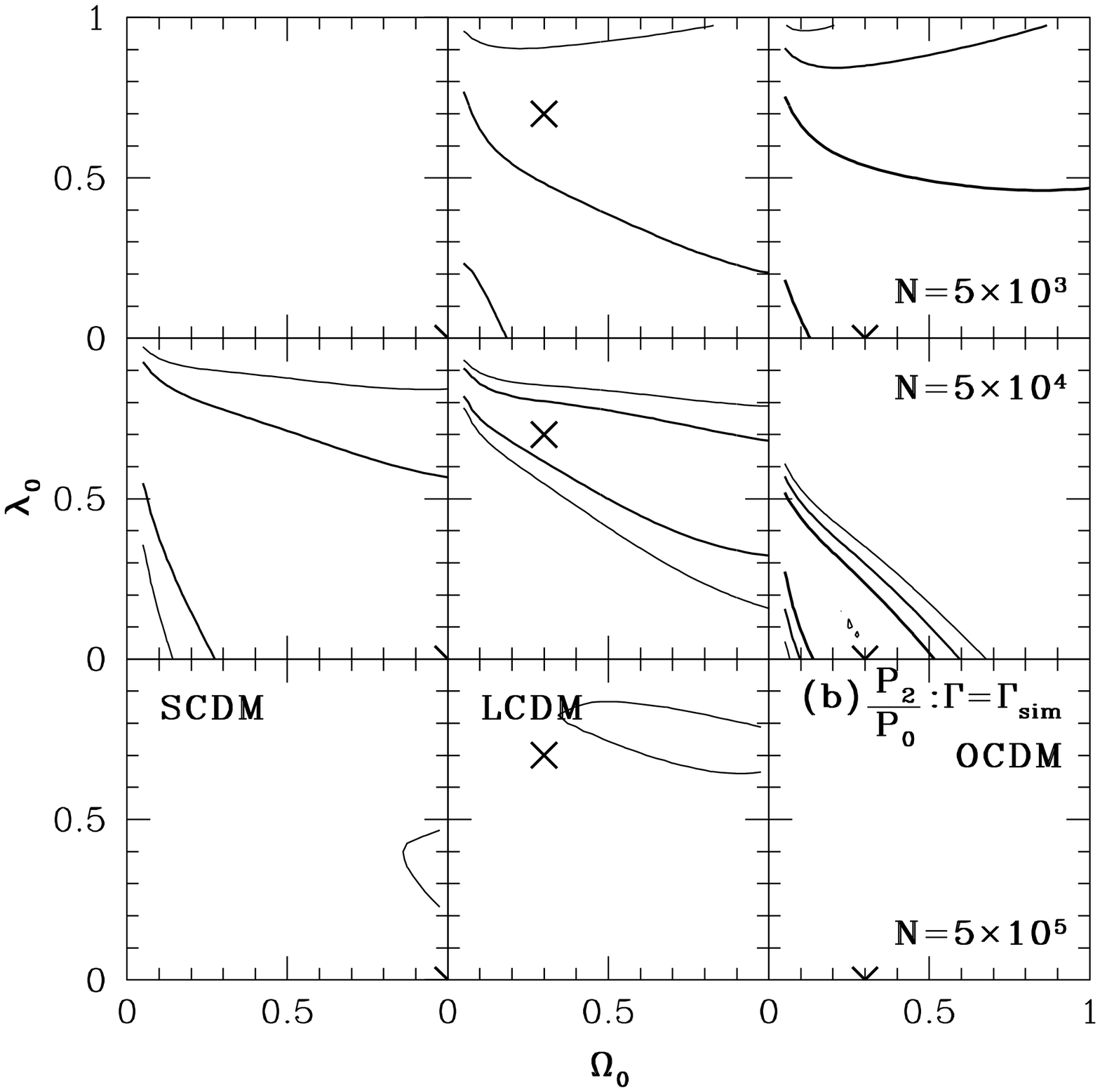}
\end{center}
\figcaption{
The same as Figure \protect\ref{fig:chi2p0p2ol} but from
  the analysis of the quadrupole to monopole ratio.
\label{fig:chi2p2ovp0ol}
}
\end{figure}

Finally we repeat the similar analysis for the quadrupole to
monopole ratio, the results of which are plotted in Figure
\ref{fig:chi2p2ovp0ol}. As anticipated the resulting constraints are
more sensitive to $\lambda_0$ compared to those presented in Figures
\ref{fig:chi2p02ol}. On the other hand, we realize that our current
modeling may not be sufficiently good to describe the ratio in
practice; note that while the amplitudes of the monopole and the
quadrupole span two orders of magnitudes, their ratio is merely around
between $-2$ and 1, and thus the accurate modeling and good data
qualities are required for the robust estimation of the cosmological
parameters. This is why some panels in Figure \ref{fig:chi2p2ovp0ol}
do not have any contours with reasonable confidence levels.

\section{Conclusions and discussion}

In this paper, we have examined the reliability and accuracy of the
theoretical modeling for the power spectrum in cosmological redshift
space using the high-resolution N-body simulations. Our main
conclusion is that the cosmological test proposed by Ballinger et al.
(1996) and Matsubara \& Suto (1996) is in fact practically useful in
constraining the cosmological parameters combined with the upcoming
SDSS QSO sample.  While an application of this methodology to the
Lyman-break galaxies is also interesting, the small number of total
objects and cosmic variance seem to be the major difficulties in
practice (Nair 1999). The results are admittedly complicated since
many different and important contaminations are necessarily involved.
In particular, our methodology is heavily dependent on the underlying
mass power spectrum, which may be reconstructed from the redshift-space
observation but with large uncertainties. Thus we explicitly assume
that the spectrum is completely specified by four free parameters,
$\Omega_0$, $\lambda_0$, $\Gamma$ and $\sigma_8$, as in the case of
CDM models.  More importantly and realistically, any model for bias
should add at least another free parameter (even in a time-independent
and scale-invariant linear model which is too idealistic).  As a
result, the estimated parameters are inevitably model-dependent.
Therefore in order to extract useful cosmological information, we have
to combine the additional constraints on those parameters from
independent cosmological consideration. Even so, the present model is
a minimal requirement to understand the clustering of high-redshift
objects properly.

The remaining important problems to improve this cosmological test
include the higher-order moment analysis, the biasing and the
light-cone effects; first, we have shown that the anisotropy of the
power spectrum in cosmological redshift space shows up already in the
monopole and quadrupole moments, $P_0^{(\CRD)}(k_s)$ and
$P_2^{(\CRD)}(k_s)$. On the other hand, this is why the resulting
constraints are only weakly dependent on $\lambda_0$, unlike the
original idea by Alcock \& Paczy\'nski (1979); note, however, that
Ballinger et al.  (1996) correctly recognized this difficulty.
Utilizing the higher-order moments will be another possibility,
although the data will be necessarily noisier.  Second, our analysis
presented above implicitly assumes to apply for the data in a narrow
redshift bin. In order to increase the number of available pairs, and
thus the statistical significance, it is crucial to take account of
the cosmological light-cone effect properly in theoretical predictions
following Yamamoto \& Suto (1999), Yamamoto, Nishioka, \& Suto (1999)
and Suto et al. (1999a). Finally, we did not allow for the possible
biasing which should further complicate the mapping of the clustering
statistics of the luminous objects to that of the underlying mass
distribution. These issues will be discussed elsewhere (Magira, Jing,
Taruya, \& Suto 1999).

\bigskip
\bigskip

We thank Takahiko Matsubara for discussion, and the referee, Alan
Heavens, for several constructive comments.  Y. P. J. gratefully
acknowledges support from a JSPS (Japan Society for the Promotion of
Science) fellowship.  Numerical computations presented were carried
out on VPP300/16R and VX/4R at ADAC (the Astronomical Data Analysis
Center) of the National Astronomical Observatory, Japan, as well as at
RESCEU (Research Center for the Early Universe, University of Tokyo)
and KEK (High Energy Accelerator Research Organization, Japan). This
research was supported in part by the Grant-in-Aid by the Ministry of
Education, Science, Sports and Culture of Japan (07CE2002) to RESCEU,
and by the Supercomputer Project (No.98-35 and No.99-52) of KEK.

\clearpage
\bigskip
\bigskip
\parskip 1pt
\baselineskip 13pt

\centerline{\bf REFERENCES}

\medskip

\def\apjpap#1;#2;#3;#4; {\pp#1, {#2}, {#3}, #4}
\def\apjbook#1;#2;#3;#4; {\pp#1, {#2} (#3: #4)}
\def\apjppt#1;#2; {\pp#1, #2}
\def\apjproc#1;#2;#3;#4;#5;#6; {\pp#1, {#2} #3, (#4: #5), #6}

\apjpap Alcock, C., \& Paczy\'nski, B. 1979;Nature;281;358;
\apjpap Ballinger, W. E., Peacock, J. A., \& Heavens, A. F. 1996;MNRAS;282;877;
\apjpap Bardeen, J. M., Bond, J. R., Kaiser, N., \& Szalay, A. S. 1985;ApJ;304;15;
\apjpap Boyle, B. J., Shanks, T., \& Peterson, B. A. 1988;MNRAS;235;935;
\apjpap Cole, S., Fisher, K. B., \& Weinberg, D. H. 1994;MNRAS;267;785;
\apjpap Cole, S., Fisher, K. B., \& Weinberg, D. H. 1995;MNRAS;275;515;
\apjpap Davis, M., \& Peebles, P. J. E. 1983;ApJ;267;465;
\apjpap Dekel, A. , \& Lahav, O. 1999;ApJ;520;24;
\apjpap de Laix, A. A., \& Starkman, G. D. 1998;MNRAS;299;977;
\apjpap Efstathiou, G., Frenk, C. S., White, S. D. M., \& Davis, M.
 1988;MNRAS;235;715;
\apjpap Fan, X., et al. 1999;AJ;118;1;
\apjpap Fry, J. N. 1996;ApJ;461;L65;
\apjpap Garnavich, P. M., et al. 1998;ApJ; 509; 74;
\apjpap Haiman, Z., \& Loeb, A. 1999;ApJ;521;L9;
\apjppt Hamilton, A. J. S. 1998; in `` The Evolving Universe. Selected
Topics on Large-Scale Structure and on the Properties of Galaxies'',
(Kluwer: Dordrecht), p.185;
\apjpap Hatton, S., \& Cole, S. 1998;MNRAS;296;10;
\apjppt Jing, Y. P. 1992;ph.D.thesis, SISSA, Trieste;
\apjpap Jing, Y. P. 1998;ApJ;503;L9;
\apjpap Jing, Y. P., \& Suto, Y. 1998;ApJ;494;L5;
\apjpap Juszkiewicz, R., Fisher, K. B., \& Szapudi, I. 1998;ApJ;504;1;
\apjpap Kaiser, N. 1984;ApJ;284;L9;
\apjpap Kaiser, N. 1987;MNRAS;227;1;
\apjpap Kitayama, T., \& Suto, Y. 1997;ApJ;490;557;
\apjpap La Franca, F., Andreani, P., \& Cristiani, S. 1998;ApJ;497;529;
\apjpap Lahav, O., Piran, T., \& Treyer, M. 1997;
MNRAS;284;499;
\apjppt Magira, H., Jing, Y. P., Taruya, \& Suto, Y. 1999;in preparation;
\apjpap Matarrese, S., Coles, P., Lucchin, F., \& Moscardini, L. 1997;
 MNRAS;286;115;
\apjpap Matsubara, T., \& Suto, Y. 1996;ApJ;470;L1;
\apjpap Matsubara, T., Suto, Y., \& Szapudi, I. 1997;ApJ;491;L1;
\apjpap Mo, H. J., \& White, S. D. M. 1996;MNRAS;282;347;
\apjpap Mo, H. J., Jing, Y. P., \& B\"orner, G. 1997;MNRAS;286;979 (MJB);
\apjpap Moscardini, L., Coles, P., Lucchin, F., \& Matarrese, S. 1998
   ;MNRAS;299;95;
\apjpap Nakamura, T. T., Matsubara, T., \& Suto, Y. 1998;ApJ;494;13;
\apjppt Nair, V. 1999;ApJ, in press (astro-ph/9904312);
\apjpap Nishioka, H., \& Yamamoto, K. 1999;ApJ;520;426;
\apjpap Peacock, J. A., \& Dodds, S. J. 1994;MNRAS;267;1020;
\apjpap Peacock, J. A., \& Dodds, S. J. 1996;MNRAS;280;L19 (PD);
\apjpap Perlmutter, S., et al. 1999;ApJ;517;565;
\apjpap Popowski, P. A., Weinberg, D. H., Ryden, B. S., 
\& Osmer, P. S. 1998;ApJ;498;11;
\apjpap Seto, N., \& Yokoyama, J. 1998;ApJ;492;421;
\apjpap Steidel, C. C., Giavalisco, M., Pettini, M., Dickinson, M., \&
Adelberger, K. L. 1996;ApJ; 462;L17;
\apjpap Steidel, C. C., Adelberger, K. L., Dickinson, M., Giavalisco,
M., Pettini, M., \& Kellogg, M. 1998;ApJ;492;428;
\apjpap Sugiyama N. 1995; ApJS; 100; 281; 
\apjpap Suto, Y. 1993;Prog.Theor.Phys.;90;1173;
\apjpap Suto, Y., Magira, H., Jing, Y. P., Matsubara, T., 
\& Yamamoto, K. 1999a;Prog.Theor.Phys.Suppl.;133;183;
\apjppt Suto, Y.,  Yamamoto, K., Kitayama, T., \& Jing, Y. P. 
1999b;ApJ, submitted (astro-ph/9907105);
\apjpap Suto, Y., \& Suginohara, T. 1991; ApJ;370;L15;
\apjpap Taruya, A., Koyama, K., \& Soda, J. 1999;ApJ;510;541;
\apjpap Treyer, M., Scharf, C., Lahav, O., Keith, J., Boldt, E., \&
Piran, T. 1998; ApJ;509;531;
\apjpap Ueda, H., Itoh, M., \& Suto, Y. 1993;ApJ;408;3;
\apjpap Wallington, S., \& Narayan, R. 1993;ApJ;403;517;
\apjpap Yamamoto, K., \& Suto, Y. 1999;ApJ;517;1;
\apjpap Yamamoto, K., Nishioka, H., \& Suto, Y. 1999;ApJ;527;
 December 20 issue, in press (astro-ph/9908006);


\vspace*{-1.0cm}

\clearpage

\begin{table}
\begin{center}
\caption{Simulation model parameters. \label{tab:modelparam}}
\medskip
\begin{tabular}{lcccccccc}
\tableline
Model & $\Omega_0$ &  $\lambda_0$  
&  $\Gamma$\tablenotemark{a} & $\sigma_8$  & $N$ & $h$\tablenotemark{b} & $A$\tablenotemark{c} & realizations \\ 
\tableline
SCDM (Standard CDM) & 1.0 & 0.0 & 0.5  & 0.6 & $256^3$ & 0.50 & 0.60 & 3  \\
LCDM (Lambda CDM) & 0.3 & 0.7 & 0.21 & 1.0 & $256^3$ & 0.70 & 0.52 & 3  \\
OCDM (Open CDM) & 0.3 & 0.0 & 0.25 & 1.0 & $256^3$ & 0.83 & 0.55 & 3  \\
\tableline 
\end{tabular}
\end{center}
\tablenotetext{a}{the shape parameter of the power spectrum.}
\tablenotetext{b}{the dimensionless Hubble constant 
{\it defined} through $h\equiv\Gamma/\Omega_0$ neglecting the
baryon density parameter $\Omega_b=0$.}
\tablenotetext{c}{the normalization factor of $\sigma_8$-$\Omega_0$ relation
determined from the X-ray cluster abundance (Kitayama \& Suto 1997).}
\end{table}


\begin{table}
\begin{center}
\caption{Pairwise peculiar velocity dispersions at $z=0$ and
$2.2$. \label{tab:fitparam}}
\medskip
\begin{tabular}{cccccc}
\tableline
Model & $z$ & $\sigma_{\scriptscriptstyle {\rm \P,MJB}}$[km/s] &
$\sigma_{\scriptscriptstyle {\rm \P,sim}}$[km/s]\tablenotemark{a}&
$\sigma_{\scriptscriptstyle {\rm \P,fit}}$[km/s]\tablenotemark{b}& 
fitting range[km/s]\tablenotemark{c}\\
\tableline
SCDM & 0   & 580 & 592 & 493 & 900 - 2900 \\
LCDM & 0   & 582 & 606 & 563 & 900 - 2900  \\
OCDM & 0   & 599 & 603 & 602 & 900 - 2900  \\
\tableline
SCDM & 2.2 & 164 & 166 & 113 & 100 - 1100  \\
LCDM & 2.2 & 381 & 379 & 273 & 900 - 2900  \\
OCDM & 2.2 & 368 & 365 & 287 & 900 - 2900  \\
\tableline
\end{tabular}
\end{center}
\tablenotetext{a}{a relative pairwise peculiar velocity dispersion directly 
evaluated from the N-body data at $r=42 h^{-1}$Mpc.}
\tablenotetext{b}{a relative pairwise peculiar velocity dispersion evaluated 
through a fit to the exponential distribution function.}
\tablenotetext{c}{a range of $v_{12}$ for the fit.}
\end{table}

\end{document}